# Optimal Selective Feedback Policies for Opportunistic Beamforming

Tharaka Samarasinghe, *Student Member, IEEE,* Hazer Inaltekin, *Member, IEEE,* and Jamie S. Evans, *Member, IEEE*

**Abstract**

This paper studies the structure of downlink sum-rate maximizing selective decentralized feedback policies for opportunistic beamforming under finite feedback constraints on the average number of mobile users feeding back. Firstly, it is shown that any sum-rate maximizing selective decentralized feedback policy must be a threshold feedback policy. This result holds for all fading channel models with continuous distribution functions. Secondly, the resulting optimum threshold selection problem is analyzed in detail. This is a non-convex optimization problem over finite dimensional Euclidean spaces. By utilizing the theory of majorization, an underlying Schur-concave structure in the sum-rate function is identified, and the sufficient conditions for the optimality of homogenous threshold feedback policies are obtained. Applications of these results are illustrated for well known fading channel models such as Rayleigh, Nakagami and Rician fading channels, along with various engineering and design insights. Rather surprisingly, it is shown that using the same threshold value at all mobile users is not always a rate-wise optimal feedback strategy, even for a network with identical mobile users experiencing statistically the same channel conditions. For the Rayleigh fading channel model, on the other hand, homogenous threshold feedback policies are proven to be rate-wise optimal if multiple orthonormal data carrying beams are used to communicate with multiple mobile users simultaneously.

This research was supported by the Australian Research Council under Grant DP-09-84862 and Grant DP-11-0102729. The material in this paper was presented in part at the International Symposium of Modeling and Optimization in Mobile Ad Hoc, and Wireless Networks, Princeton, USA, May 2011, and at the IEEE International Symposium on Information Theory, St. Petersburg, Russia, August 2011.

The authors are with the Department of Electrical and Electronic Engineering, University of Melbourne, Parkville, VIC 3010, Australia (e-mails: ts@unimelb.edu.au, hazeri@unimelb.edu.au, jse@unimelb.edu.au).

# I. INTRODUCTION

## A. Background and Motivation

A key design challenge in fourth generation (4G) wireless networks is to achieve data rates as high as 1 Gbit/s for low mobility and 100 Mbit/s for high mobility [1]. Multiple-input multiple-output (MIMO) technology marks a paradigm change from scalar communication to the *vector* one, and has now become an integral part of 4G wireless networks to accomplish such high data rate targets. Benefits include *power gain*, *diversity gain* and *degrees-of-freedom gain* (or, alternatively called multiplexing gain) [2]–[4], to name a few. When there is a multitude of mobile users (MU) in the communication environment, which is usually the case in a network setting, random beamforming techniques can also be used to utilize *multiuser diversity gain* [5].

Some of these gains can be harvested without requiring any knowledge of wireless channel states such as diversity gain, but some others can be exploited effectively only through some form of channel state information (CSI) at the base station (BS) [6], [7]. In such communication instances necessitating the use of CSI for adaptive signaling, feedback is an important means to convey required information from MUs to the BS. This paper studies rate-wise optimal *selective* feedback policies for vector broadcast channels, and establishes the structure of such policies under finite feedback constraints.

We consider the classical opportunistic communication along multiple orthonormal beams. The focus is on the total downlink communication rate[1], and the BS is provided only with *partial* CSI (*i.e.*, downlink $\mathrm{SINR}$ values) for scheduling such as in the IS-856 standard. Hence, the (full CSI) sum-rate capacity achieving dirty paper precoding [8]–[12], or any other transmit beamforming strategy requiring full CSI to this end, is automatically disallowed. The wireless channels, and therefore the attained signal-to-interference-plus-noise ratios ($\mathrm{SINR}$) by different users on different beams, change over time. The BS selects the best user (with the highest $\mathrm{SINR}$) per beam to maximize the sum-rate at the downlink.

This is the opportunistic beamforming (OBF) approach utilizing multiuser diversity and varying channel conditions to extract all degrees-of-freedom available for the downlink communication (provided by the use of multiple transmit antennas) as well as to deliver improved power gains [5], [13]. Indeed, it achieves the same full CSI sum-rate capacity to a first order for large numbers of MUs in the network [13]. However, for large numbers of MUs, the OBF approach still requires large amounts of data to be fed back, which is an onerous requirement on the uplink feedback channel. What is needed is a selective decentralized feedback policy that will only choose a small subset of MUs to be multiplexed on the uplink feedback channel. In this case, the downlink sum-rate is certainly a function of the feedback policy selecting MUs. We ask: What is the structure of the sum-rate maximizing selective decentralized feedback policies, and how does the resulting sum-rate compare to the sum-rate without any user selection?

[1]Unless otherwise stated, we use the term *rate* (or, its derivatives such as sum-rate, total rate, aggregate communication rate) to always refer to the *ergodic rate* obtained by averaging over many fading states.

### *B. Contributions*

Our main findings can be summarized as follows. We first show that any sum-rate maximizing selective decentralized feedback policy for a given constraint on the average number of users feeding back must be a *threshold feedback policy* in which each MU, independently from others, decides to feed back or not by comparing its $\mathsf{SINR}$ values with a predetermined threshold value. Different MUs are allowed to have different thresholds if such heterogeneity in thresholds maximizes the total downlink rate. This thresholding optimality result does not depend on the particular statistical model of the wireless channel as long as the resulting $\mathsf{SINR}$ distribution is continuous, which holds for most common fading models such as Rayleigh, Rician and Nakagami fading. It also possesses a *stability* property from a game theoretic point of view as explained in Section IV.

These findings provide an analytical justification for the use of threshold feedback policies in practical systems, and strengthen previous work on thresholding as an appropriate selective feedback scheme, *e.g.*, see [13]–[18]. They also form a basis for the optimum threshold selection problem analyzed in Section V. To some extent, our thresholding optimality result is intuitive and expected. It is even known to hold in the limit without feedback constraints for richly scattered Rayleigh fading environments [16], [18]. However, its proof in our case is not straightforward, and requires a careful analysis of rate gain and loss events due to coupling effects, induced by finite feedback constraints, of MUs' individual feedback rules on the sum-rate function.

Having established the optimality of threshold feedback policies, we now face the *optimum threshold selection problem* to further maximize the downlink sum-rate. This optimization problem is over the familiar finite dimensional Euclidean spaces, but it turns out that the objective sum-rate function is not necessarily convex as a function of MUs' threshold values. Thus, we resort to the theory of majorization [19], and solve the optimum threshold selection problem by identifying an underlying *Schur-concave* structure in the sum-rate function. In particular, we obtain sufficient conditions for the Schur-concavity of the sum-rate, and therefore for the rate optimality of *homogenous* threshold feedback policies in which all MUs use the same threshold for their feedback decisions. These conditions are provided for general fading models under some mild conditions on the resulting $\mathsf{SINR}$ distribution, which are satisfied by most common fading models such as Rayleigh, Rician and Nakagami fading.

A naive but intuitive approach to maximize the total downlink communication rate for a network with identical MUs experiencing statistically the same channel conditions is to use a homogenous threshold feedback policy satisfying feedback constraints. Rather surprisingly, our results reveal that this intuition does not always work here. We provide a simple counterexample in which only a single beam is used for the downlink communication with two MUs in a Rayleigh fading environment. In the high signal-to-noise-ratio (SNR) regime, necessary conditions for the Schur-concavity of the sum-rate are violated, and it becomes strictly suboptimal to use the same threshold value to mediate MUs' feedback decisions. Indeed, we prefer one MU over the other one by assigning a small threshold for this MU to minimize the feedback outage event probability, *i.e.,* the probability that none of the MUs feeds back. On the other hand, we show that the sum-rate is a Schur-concave function when the SNR is low, and therefore the homogenous threshold feedback policy satisfying feedback constraints with equality is the optimum policy to maximize the sum-rate in the low SNR regime. To put it in another way, we trade the power gain (due to multiuser diversity) for the degrees-of-freedom gain (due to minimum outage communication) in the high SNR regime, whereas the degrees-of-freedom gain is traded for the power gain in the low SNR regime. An extensive numerical study utilizing our sufficient conditions is also performed to illustrate optimality and sub-optimality regions for the homogenous threshold feedback policies for fading models other than Rayleigh fading such as Rician and Nakagami fading.

On the more positive side, we show that the sum-rate is always a Schur-concave function for all values of SNR when *two or more* orthonormal beams are used to simultaneously communicate with multiple MUs located in a Rayleigh fading environment. In this case, the downlink communication becomes interference limited, rather than noise limited, due to inter-beam interference, and therefore the behavior of the optimum threshold feedback policy becomes unchanged for all SNR values: Use the same threshold for all MUs such that the feedback constraint is satisfied with equality. For this fading scenario, the difference between communication rates achieved with and without user selection is also illustrated. In particular, when the threshold values are optimally set for large user populations, there is almost no rate loss if the average number of MUs feeding back per beam is around five. From a practical point of view, this signifies a significant reduction in the feedback load without noticeable performance loss, and provides an important cross-layer design parameter for the higher MAC layer for multiplexing MUs on the uplink to feed back.

The remainder of the paper is organized as follows. In Section II, we compare and contrast our results with the relevant previous work. Section III describes the system model, provides basic concepts and definitions to be used in the rest of the paper, and formulates the problem of finding the optimal feedback policy maximizing the aggregate communication rate under finite feedback constraints as a function optimization problem. In Section IV, we show that this function optimization problem can be reduced to a finite dimensional but non-convex optimum threshold selection problem by establishing the optimality of threshold feedback policies. In Section V, we solve the optimum threshold selection problem by using the theory of majorization. In particular, sufficient conditions for the Schur-concavity of the sum-rate function are derived. Section VI presents an extensive numerical and simulation study to illustrate the applications of these results to familiar fading models along with various engineering and design insights. Section VII concludes the paper.

## II. Related Work

Feedback load reduction techniques for adaptive signaling in wireless communication networks have been a key area of research for more than a decade, especially with the advent of MIMO technology, *e.g.*, see [7] and the references therein for an overview of feedback load reduction techniques in wireless communication systems. Among many promising approaches proposed over the last decade, OBF (*a.k.a.*, opportunistic beamforming) has attracted considerable attention and research effort since its inception by Viswanath et al. in [5]. It is a practical way of reducing feedback requirements for vector broadcast channels, yet still achieves the full CSI sum-rate capacity at the downlink to a first order [13]. In this paper, we are also motivated by such opportunistic communication and beamforming techniques, and focus on the downlink sum-rate maximization under finite feedback constraints on the feedback uplink.

Capacity scaling laws attained by OBF were first obtained by Shariff and Hassibi in [13]. Among many other results, they, in particular, showed that if an opportunistic scheduling algorithm is used to harvest multiuser diversity gains, the downlink throughput scales optimally like $M \log \log n$, where $M$ is the number of transmit antennas at the BS, and $n$ is the number of MUs in the system. In [20], the authors built upon [13] to derive tighter expressions for the downlink sum-rate scaling for OBF. Unlike these works, the results derived for the structure and optimization of the downlink sum-rate in this paper are correct for any number (small and large) of MUs in the network. In addition, the sum-rate maximization problem addressed in this paper does not appear in [13] and [20].

Without any user selection, the number of MUs feeding back grows linearly with the total number of MUs in the system to achieve double logarithmic growth in the downlink sum-rate. Threshold feedback policies are frequently used to alleviate such an excessive feedback requirement on the uplink [13]–[18]. In [14], the authors proposed to use a common threshold level to arbitrate MUs' feedback decisions for scalar channels. They showed that this approach has the potential to significantly reduce the total feedback load on the uplink while maintaining *almost* the same sum-rate performance at the downlink. In [15], the authors extended the feedback scheme proposed in [14] by using multiple threshold levels. This paper differs from [14] and [15] in three important aspects. Firstly, we provide an analytical justification for why threshold feedback policies are right choice for user selection, *e.g.*, see Section IV for details. Secondly, we pose an optimum threshold selection problem in which we search for the optimum assignment of thresholds to MUs. We show that using the same threshold value for all MUs is not always optimum even if all MUs experience statistically the *same* channel conditions. Finally, our results are given for more general vector broadcast channels.

In [13] and [17], the authors used a constant threshold level, the same for all MUs and independent of the number of MUs, to reduce the total feedback load for vector broadcast channels within the OBF framework. Such a constant thresholding scheme cannot eliminate the linear growth in the average number of MUs feeding back. In [16], it was shown that it is enough to have only $O(\log n)$ MUs feeding back to achieve the same downlink sum-rate scaling in [13] by varying the common threshold level with the total number of MUs in the system. This result was extended in [18] by showing that $O\left((\log n)^{\epsilon}\right)$, $\epsilon \in (0,1)$, MUs are enough to achieve the same downlink sum-rate scaling in [13]. It is *almost as if* constant feedback load is enough to maintain optimum sum-rate scaling but not exactly. In contrast to these previous works, we focus on more stringent but practical *constant* feedback requirements in this paper. The sum-rate maximization framework introduced here does not exist in these papers, either. Finally, these previous works only focused on the asymptotic sum-rate scaling behavior, whereas our results are correct for any finite number of MUs in the network.

An important issue associated with OBF is its applicability to finite networks. [21]–[24] propose various methods for optimizing OBF for smaller sets of MUs. [21] and [22] propose algorithms to select a target group of MUs, and then to request perfect CSI only from the selected set of MUs to facilitate more efficient beamforming schemes. [23] and [24] show how feedback aggregation and multiple beamforming vectors can be utilized to fine-tune OBF, respectively. Similar to these works, we also focus on finite networks in this paper. However, our problem set-up and motivation are much different than those in [21]–[24]. Here, we are interested in the structure of feedback policies maximizing the downlink sum-rate given the constraints on the average number of MUs to be multiplexed on the uplink for feedback.

Fairness is also among the important topics for OBF. Proportionally fair algorithm proposed in [5] ensures long-term fairness among MUs in terms of average data rates achieved. Although indirectly, this paper reveals an interesting and somewhat counterintuitive observation in regards to fairness in the OBF framework. Even for a network with statistically identical MUs, we show that it may become more favorable to treat MUs unequally, *i.e.*, to prefer one group of MUs over others by allocating the wireless channel to them more frequently, to maximize the downlink sum-rate. We also obtain various sufficient conditions on wireless channel statistics under which fairness is automatically achieved, *i.e.*, all MUs are given equal chances to feed back and to access the channel. A detailed discussion on this account is provided in Sections V and VI.

Related work also includes [25]–[28]. In [25]–[27], CSI parameters were quantized to reduce the feedback load for OBF. This approach cannot eliminate the linear feedback load growth alone, but it leads to further feedback reductions when combined with a user selection protocol. In this paper, we solve the optimum threshold selection problem offline under statistical information about wireless channels. Once the thresholds are optimally assigned for user selection, it is an added design choice how to quantize $\mathrm{SINR}$ parameters, and the resulting performance analysis requires further investigation, which we do not address in this paper.

In [28], the authors focused on exploiting multiuser diversity in a distributed manner for scalar multiple access channels by means of thresholds. Their MAC layer consisted of a collision channel model, and the thresholds were chosen to be the same for all MUs with identical channel statistics. Although we focus on the *dual* vector broadcast channels in this paper without any attention on the multiple access feedback uplink, our results have some ramifications for the MAC problem studied in [28]. First of all, our homogenous threshold optimality results imply that using different threshold levels for different MUs with identical channel conditions may further improve the data rates reported in [28]. Secondly, they provide a cross-layer design parameter for the number of MUs to be multiplexed on the uplink (for feedback) without any noticeable performance degradation at the downlink.

## III. System Model and Problem Formulation

We consider a multi-antenna single cell vector broadcast channel. There are $n$ MUs in the cell. The BS has $N_t$ transmit antennas, and each MU is equipped with a single receive antenna. The channel gains between the receive antenna of the $i$th MU and the transmit antennas of the BS are given by $\boldsymbol{h}_i = (h_{1,i}, \ldots, h_{N_t,i})^\top$, where $h_{k,i}$ is the channel gain between the $k$th transmit antenna at the BS and the receive antenna at the $i$th MU. We assume that $h_{k,i}$, $k = 1, \ldots, N_t$ and $i = 1, \ldots, n$, are independent and identically distributed (i.i.d.) random variables. In addition, we assume a quasi-static block fading model, in which channel gains are constant through a coherence time interval, and change from one coherence period to another independently according to a common fading distribution. For the sake of notational simplicity, we drop the time index here in the channel model, and also later in the representation of transmitted and received signals.

Our signal model is similar to the one in [13]. The BS transmits $M$, $M \leq N_t$, different data streams intended for $M$ different MUs. The symbols of the $k$th stream are represented by $s_k$. They are chosen from the capacity achieving unit power (complex) Gaussian codebooks, and are sent along the directions of $M$ orthonormal beamforming vectors $\left\{ \boldsymbol{b}_k = (b_{1,k}, \ldots, b_{N_t,k})^\top \right\}_{k=1}^{M}$. These beamforming vectors can be either deterministic, or randomly generated and updated periodically. The overall transmitted signal from the BS is given by

$$\boldsymbol{s} = \sqrt{\rho} \sum_{k=1}^{M} \boldsymbol{b}_k s_k, \tag{1}$$

where $\rho$ is the transmit power per beam. The signal received by the $i$th MU is equal to

$$Y_i = \sqrt{\rho} \sum_{k=1}^{M} \boldsymbol{h}_i^\top \boldsymbol{b}_k s_k + Z_i, \tag{2}$$

where $Z_i$ is the unit power (complex) Gaussian background noise. With these normalized parameter selections, $\rho$ also signifies the SNR per beam as in [13]. Let $\gamma_{m,i}$ be the SINR value corresponding to the $m$th beam at the $i$th MU. Then, it is given by

$$\gamma_{m,i} = \frac{|\boldsymbol{h}_i^\top \boldsymbol{b}_m|^2}{\rho^{-1} + \sum_{k=1, k \neq m}^{M} |\boldsymbol{h}_i^\top \boldsymbol{b}_k|^2}. \tag{3}$$

Let $\boldsymbol{\gamma}_i = (\gamma_{1,i}, \ldots, \gamma_{M,i})^\top \in \mathbb{R}_+^M$ represent the SINR vector at MU $i$. Beams are statistically identical, and the elements of $\boldsymbol{\gamma}_i$ are identically distributed for all $i \in \mathcal{N}$ with a common marginal distribution $F$, where $\mathcal{N} = \{1, \ldots, n\}$. However, SINR values at a particular MU are dependent random variables, *i.e.*, see (3). We will assume that $F$ is continuous, and has the density $f$ with support $\mathbb{R}_+$, which are true for many fading models including Rayleigh, Rician and Nakagami fading. Similar assumptions on the fading distribution also exist in [28]. For the ease of notation, if $M = 1$, we will use $\gamma_i$ to denote the SINR of MU $i$ on this single beam. $\boldsymbol{\Gamma} = [\boldsymbol{\gamma}_1, \ldots, \boldsymbol{\gamma}_n] \in \mathbb{R}_+^{M \times n}$ is the system-wide $M$-by-$n$ SINR matrix that contains the SINR vectors of all MUs in the system.

If the BS has perfect knowledge of $\boldsymbol{\Gamma}$, the aggregate communication rate can be maximized by choosing the best MU with the highest SINR on each beam. However, this necessitates excessive amount of feedback and information exchange between the BS and MUs. Therefore, we focus on the sum-rate maximization under finite feedback constraints, where MUs feed back according to a predefined selective feedback policy as defined below.

Let $\gamma_i^\star = \max_{1 \leq k \leq M} \gamma_{k,i}$ be the maximum SINR value at MU $i$, and let $b_i^\star = \arg\max_{1 \leq k \leq M} \gamma_{k,i}$ be the index of the best beam achieving $\gamma_i^\star$. Let also $\mathcal{M} = \{1, \ldots, M\}$. Using these notations, we formally define a feedback policy as follows.

*Definition 1:* A feedback policy $\boldsymbol{\mathcal{F}} : \mathbb{R}_+^{M \times n} \mapsto \{\Omega \bigcup \{\emptyset\}\}^n$ is an $\{\Omega \bigcup \{\emptyset\}\}^n$-valued function $\boldsymbol{\mathcal{F}} = (\mathcal{F}_1, \ldots, \mathcal{F}_n)^\top$, where $\mathcal{F}_i : \mathbb{R}_+^{M \times n} \mapsto \Omega \bigcup \{\emptyset\}$ is the feedback rule of MU $i$, $\Omega$ is the set of all feedback packets and $\emptyset$ represents the no-feedback state. We call $\boldsymbol{\mathcal{F}}$ a *general decentralized feedback policy* if $\mathcal{F}_i$ is only a function of $\boldsymbol{\gamma}_i$ for all $i \in \mathcal{N}$. We call it a *homogenous general decentralized feedback policy* if it is decentralized and all MUs use the same feedback rule. Finally, we call it a *maximum* SINR *decentralized feedback policy*, if $\mathcal{F}_i(\boldsymbol{\gamma}_i)$ is only a function of $\gamma_i^\star$ and $b_i^\star$, and produces a feedback packet containing $\gamma_i^\star$ as the sole SINR information on a positive feedback decision, and otherwise produces $\emptyset$.

Intuitively, a feedback policy determines whether a MU will feed back or not. Upon a positive feedback decision, it generates a feedback packet containing SINR values at selected beams (along with other information to be contained in the packet header), and sends it to the BS for central processing. When it is clear from the context, we will omit the term "general". We will index system-wide feedback policies by superscripts such as $\boldsymbol{\mathcal{F}}^i$, and individual feedback rules by subscripts such as $\mathcal{F}_i$. We use the term "policy" to refer to system-wide feedback rules, whereas the term "rule" is used to refer to individual feedback rules. The definitions given for system-wide feedback policies extend to individual feedback rules in an obvious way when possible. We assume that there is no cooperation between different MUs, which is true for most practical systems, therefore we can narrow down our study to decentralized feedback policies for the system in consideration.

Furthermore, we will focus our attention on beam symmetric feedback policies since beams are assumed to be statistically identical. We formally define beam symmetric policies as follows.

*Definition 2:* Let $\Pi : \mathbb{R}^M \mapsto \mathbb{R}^M$ be a permutation mapping, *i.e.,* $\Pi(\boldsymbol{\gamma}) = \left(\gamma_{\pi(1)}, \ldots, \gamma_{\pi(M)}\right)^\top$ for some one-to-one $\pi : \mathcal{M} \mapsto \mathcal{M}$. For $\boldsymbol{\Gamma} \in \mathbb{R}^{M \times n}$, let $\Pi(\boldsymbol{\Gamma}) = [\Pi(\boldsymbol{\gamma_1}), \ldots, \Pi(\boldsymbol{\gamma_n})]$. If $\mathcal{I}_i$ is the set of beam indexes selected by $\mathcal{F}_i(\boldsymbol{\Gamma})$, and $\pi(\mathcal{I}_i)$ is the set of beam indexes selected by $\mathcal{F}_i(\Pi(\boldsymbol{\Gamma}))$ for all $i \in \mathcal{N}$, we say $\boldsymbol{\mathcal{F}}$ is a *beam symmetric* feedback policy.

This symmetry assumption is just for the sake of notational simplicity, and the same techniques can be generalized to beam asymmetric policies by allowing different feedback policies for different beams at MUs. We let $\Xi$ denote the set of all beam symmetric decentralized feedback policies. When it is clear from the context, we will also omit the term "beam symmetric".

Given a feedback policy $\boldsymbol{\mathcal{F}}$, we have a random set of MUs $\mathcal{G}_m(\boldsymbol{\mathcal{F}}(\boldsymbol{\Gamma}))$ requesting beam $m \in \mathcal{M}$. When $\mathcal{G}_m(\boldsymbol{\mathcal{F}}(\boldsymbol{\Gamma}))$ is a non-empty set at a given fading state, the BS selects the MU with the highest SINR in this set to maximize the instantaneous communication rate in the direction of beam $m$. If $\mathcal{G}_m(\boldsymbol{\mathcal{F}}(\mathbf{\Gamma}))$ is an empty set, we say a *feedback outage event* occurs at beam $m$, and zero rate is achieved at this beam.[2]

Then, the downlink ergodic sum-rate achieved under the feedback policy $\boldsymbol{\mathcal{F}}$ is given by

$$R(\boldsymbol{\mathcal{F}}) = \mathsf{E}_{\mathbf{\Gamma}}\left[r(\boldsymbol{\mathcal{F}}, \mathbf{\Gamma})\right] = \mathsf{E}_{\mathbf{\Gamma}}\left[\sum_{m=1}^{M} \log\left(1 + \max_{i \in \mathcal{G}_m(\boldsymbol{\mathcal{F}}(\mathbf{\Gamma}))} \gamma_{m,i}\right)\right], \tag{4}$$

where $r(\boldsymbol{\mathcal{F}}, \mathbf{\Gamma})$ is the *instantaneous* sum-rate achieved under the feedback policy $\boldsymbol{\mathcal{F}}$, expectation is taken over the random SINR matrices, and the result of the maximum operation is zero when $\mathcal{G}_m(\boldsymbol{\mathcal{F}}(\mathbf{\Gamma}))$ is an empty set. $r^m(\boldsymbol{\mathcal{F}}, \mathbf{\Gamma})$ and $R^m(\boldsymbol{\mathcal{F}})$ denote the instantaneous sum-rate and the ergodic sum-rate on beam $m$, respectively. Note that $r^m(\boldsymbol{\mathcal{F}}, \mathbf{\Gamma}) = \log\left(1 + \max_{i \in \mathcal{G}_m(\boldsymbol{\mathcal{F}}(\mathbf{\Gamma}))} \gamma_{m,i}\right)$, and $R^m(\boldsymbol{\mathcal{F}}) = \mathsf{E}_{\mathbf{\Gamma}}\left[r^m(\boldsymbol{\mathcal{F}}, \mathbf{\Gamma})\right]$. Also, the sum-rate achieved on an event $\mathcal{A}$ under $\boldsymbol{\mathcal{F}}$ is written as $R(\boldsymbol{\mathcal{F}}, \mathcal{A}) = \mathsf{E}_{\mathbf{\Gamma}}\left[r(\boldsymbol{\mathcal{F}}, \mathbf{\Gamma}) \mathbf{1}_{\mathcal{A}}\right]$, and conditioned on an event $\mathcal{A}$ (or, a random variable), we define the *conditional* sum-rate as $R(\boldsymbol{\mathcal{F}}|\mathcal{A}) = \mathsf{E}_{\mathbf{\Gamma}}\left[r(\boldsymbol{\mathcal{F}}, \mathbf{\Gamma})|\mathcal{A}\right]$. We will use $R(\boldsymbol{\mathcal{F}})$ as the performance measure of a given feedback policy along the rate dimension.

Given a feedback policy $\boldsymbol{\mathcal{F}}$, we will use the average number of MUs feeding back per beam $\Lambda(\boldsymbol{\mathcal{F}})$ to measure the performance of $\boldsymbol{\mathcal{F}}$ along the feedback dimension. $\Lambda(\boldsymbol{\mathcal{F}})$ can be written as $\Lambda(\boldsymbol{\mathcal{F}}) = \sum_{i=1}^{n} p_i$, where $p_i = \mathsf{Pr}\{\mathcal{F}_i(\mathbf{\Gamma}) \text{ selects beam } 1\}$ since $\boldsymbol{\mathcal{F}}$ is beam symmetric. We are interested in maximizing the ergodic sum-rate under finite feedback constraints, and the resulting rate maximization problem can be written as

$$\begin{array}{ll} \underset{\boldsymbol{\mathcal{F}} \in \Xi}{\text{maximize}} & R(\boldsymbol{\mathcal{F}}) \\ \text{subject to} & \Lambda(\boldsymbol{\mathcal{F}}) \leq \lambda \end{array}, \tag{5}$$

*i.e.*, find the optimal feedback policy maximizing the aggregate communication rate subject to feedback constraint $\lambda$. This optimization problem is over function spaces [29], and the objective function is not necessarily convex. Firstly, we will reduce the search for optimal feedback policies to an optimal threshold selection problem over finite dimensional Euclidean spaces by proving rate-wise optimality of threshold feedback policies. Then, we will make use of an underlying Schur-concave structure in the objective function to solve the resulting optimal threshold selection problem. The next section establishes the optimality of threshold feedback policies.

[2]Note that the BS does not have access to any CSI on the feedback outage event. Without any CSI, reliable communication is still possible if we can average over very large time-scales for all MUs. The extra rate term to be added to (4) in this case would not affect our analysis in remainder of the paper, and therefore is omitted for simplicity.

## IV. Optimality of Threshold Feedback Policies

In this section, we show that the solution of the optimization problem posed in (5) must be a threshold feedback policy. We start our analysis by formally defining threshold feedback policies.

*Definition 3:* We say $\boldsymbol{\mathcal{T}} = (\mathcal{T}_1, \ldots, \mathcal{T}_n)^\top$ is a *general threshold feedback policy* (GTFP) if, for all $i \in \mathcal{N}$, there is a threshold $\tau_i$ such that $\mathcal{T}_i(\boldsymbol{\gamma}_i)$ generates a feedback packet containing SINR values $\{\gamma_{k,i}\}_{k \in \mathcal{I}_i}$ if and only if $\gamma_{k,i} \geq \tau_i$ for all $k \in \mathcal{I}_i \subseteq \mathcal{M}$. We call it a *homogenous general threshold feedback policy* if all MUs use the same threshold $\tau$, *i.e.,* $\tau_i = \tau$ for all $i \in \mathcal{N}$.

We note that a MU can be allocated to multiple beams according to Definition 3. Another class of threshold feedback policies are the feedback policies limiting each MU to request only the beam with the highest $\mathrm{SINR}$, *e.g.,* see [13], [16]–[18]. We call this class of feedback policies maximum $\mathrm{SINR}$ threshold feedback policies, and formally define them as follows.

*Definition 4:* $\boldsymbol{\mathcal{T}} = (\mathcal{T}_1, \ldots, \mathcal{T}_n)^\top$ is a *maximum* $\mathrm{SINR}$ *threshold feedback policy* (MTFP) if, for all $i \in \mathcal{N}$, there is a threshold $\tau_i$ such that $\mathcal{T}_i(\boldsymbol{\gamma}_i)$ produces a feedback packet requesting beam $k$ and containing $\gamma_{k,i}$ as the sole $\mathrm{SINR}$ information if and only if $b_i^\star = k$ and $\gamma_i^\star \geq \tau_i$.

For a given set of threshold values, it is not hard to see that the GTFP (corresponding to these threshold values) always achieves a rate at least as good as the rate achieved by the MTFP (corresponding to the same threshold values) because MUs request all the beams with $\mathrm{SINR}$ values above their thresholds under the GTFP, which includes the best beam with the highest $\mathrm{SINR}$. Since maximum $\mathrm{SINR}$ values are also fed back by GTFPs, they can be considered more general than MTFPs. Moreover, as shown later in Lemma 3, a GTFP reduces to an MTFP if threshold values of all MUs are greater than one. In this section, we will first prove that GTFPs form a rate-wise optimal subset of general decentralized feedback policies, and then obtain a similar result for MTFPs.

### A. Optimality of General Threshold Feedback Policies

It is enough to focus only on the first beam since $R(\boldsymbol{\mathcal{F}})$ can be written as

$$R(\boldsymbol{\mathcal{F}}) = M \mathsf{E}_{\boldsymbol{\Gamma}} \left[ \log \left( 1 + \max_{i \in \mathcal{G}_1(\boldsymbol{\mathcal{F}}(\boldsymbol{\Gamma}))} \gamma_{1,i} \right) \right] \tag{6}$$

under our assumptions in Section III.

For our proofs, we will define various sets whose elements lie in various spaces including $\mathbb{R}_+^M$ and $\mathbb{R}_+^{M \times n}$. Therefore, paying attention to the space in which the elements of a set lie will facilitate exposition in the rest of the paper.

For a given beam symmetric general decentralized feedback policy $\boldsymbol{\mathcal{F}} = (\mathcal{F}_1, \ldots, \mathcal{F}_n)^\top$, we let $FB_i = \left\{\boldsymbol{\gamma}_i \in \mathbb{R}_+^M : \mathcal{F}_i(\boldsymbol{\gamma}_i) \text{ selects beam } 1\right\}$ for all $i \in \mathcal{N}$. Given $\boldsymbol{\mathcal{F}}$, we construct a GTFP $\boldsymbol{\mathcal{T}}$ by choosing $\tau_i$ as $\Pr\{\gamma_{1,i} \geq \tau_i\} = \Pr\{\boldsymbol{\gamma}_i \in FB_i\}$ for all $i \in \mathcal{N}$. This construction is feasible since $\gamma_{1,i}$ is assumed to have a continuous distribution function. Such a selection of $\boldsymbol{\mathcal{T}}$ leads to a fair comparison between $\boldsymbol{\mathcal{F}}$ and $\boldsymbol{\mathcal{T}}$ since $\Lambda(\boldsymbol{\mathcal{F}}) = \Lambda(\boldsymbol{\mathcal{T}})$. We divide $FB_i$ into two disjoint sets $\mathcal{S}_i^L = \left\{\boldsymbol{\gamma}_i \in \mathbb{R}_+^M : \boldsymbol{\gamma}_i \in FB_i \ \& \ \gamma_{1,i} < \tau_i\right\}$, and $\mathcal{S}_i^R = \left\{\boldsymbol{\gamma}_i \in \mathbb{R}_+^M : \boldsymbol{\gamma}_i \in FB_i \ \& \ \gamma_{1,i} \geq \tau_i\right\}$. Finally, we let $\bar{\mathcal{S}}_i^R = \left\{\boldsymbol{\gamma}_i \in \mathbb{R}_+^M : \boldsymbol{\gamma}_i \notin FB_i \ \& \ \gamma_{1,i} \geq \tau_i\right\}$. We will use these sets to show $R(\boldsymbol{\mathcal{T}}) \geq R(\boldsymbol{\mathcal{F}})$.

The proof is simple for a single user single beam communication scenario. For a particular realization of the SINR value $\gamma_1$, the same instantaneous rate is achieved by both feedback policies if they result in the same feedback decision. On the other hand, the achieved instantaneous rate will be different if only one of the policies results in a positive feedback decision. This happens either when $\gamma_1 \in \mathcal{S}_1^L$, in which case only $\boldsymbol{\mathcal{F}}$ leads to a positive feedback decision, or when $\gamma_1 \in \bar{\mathcal{S}}_1^R$, in which case only $\boldsymbol{\mathcal{T}}$ leads to a positive feedback decision. The worst case SINR on the event $\gamma_1 \in \bar{\mathcal{S}}_1^R$ is greater than the threshold value $\tau_1$, and the best case SINR achieved by the MU on the event $\gamma_1 \in \mathcal{S}_1^L$ is less than $\tau_1$. Therefore, the rates achieved by $\boldsymbol{\mathcal{F}}$ and $\boldsymbol{\mathcal{T}}$ can be upper and lower bounded, respectively, to show that $R(\boldsymbol{\mathcal{T}}) \geq R(\boldsymbol{\mathcal{F}})$.

The proof for the multiuser scenario hinges on the same principles above but it is not straightforward due to coupling effects of individual feedback rules on the aggregate rate expression. Part of the complexity to deal with these effects arises from the heterogeneous nature of the feedback rules. For example, consider a two-user single beam communication scenario. Let $\boldsymbol{\mathcal{F}} = (\mathcal{F}_1, \mathcal{F}_2)$ be a general decentralized feedback policy, and $\boldsymbol{\mathcal{T}} = (\mathcal{T}_1, \mathcal{T}_2)$ be the corresponding general threshold feedback policy as constructed above. Consider the event $\mathcal{A}$ in which $\gamma_1 \in \bar{\mathcal{S}}_1^R$ and $\gamma_2 \in \mathcal{S}_2^L$. On this event, $\boldsymbol{\mathcal{F}}$ schedules MU 2, whereas $\boldsymbol{\mathcal{T}}$ schedules MU 1. If $\tau_2 > \tau_1$, we can envisage cases in which both $\gamma_2 > \gamma_1$ and $\gamma_1 > \gamma_2$ can happen with positive probability on $\mathcal{A}$. For example, we can represent the sets of interest defined earlier on the real line in this case (*i.e.*, $M = 1$), and Figs. 1(a) and 1(b) show example realizations of $\gamma_1$ and $\gamma_2$ for which $r(\boldsymbol{\mathcal{T}}, \mathbf{\Gamma}) < r(\boldsymbol{\mathcal{F}}, \mathbf{\Gamma})$ and $r(\boldsymbol{\mathcal{T}}, \mathbf{\Gamma}) > r(\boldsymbol{\mathcal{F}}, \mathbf{\Gamma})$, respectively. Therefore, average sum-rates cannot be bound easily to determine which feedback policy achieves higher expected rate on $\mathcal{A}$. The same arguments continue to hold for other events, and the problem complexity is further magnified with increasing numbers of MUs. To overcome these issues, we will prove a more general result indicating that the best strategy for a MU is to always use a threshold feedback policy whatever the feedback policies of other MUs are.

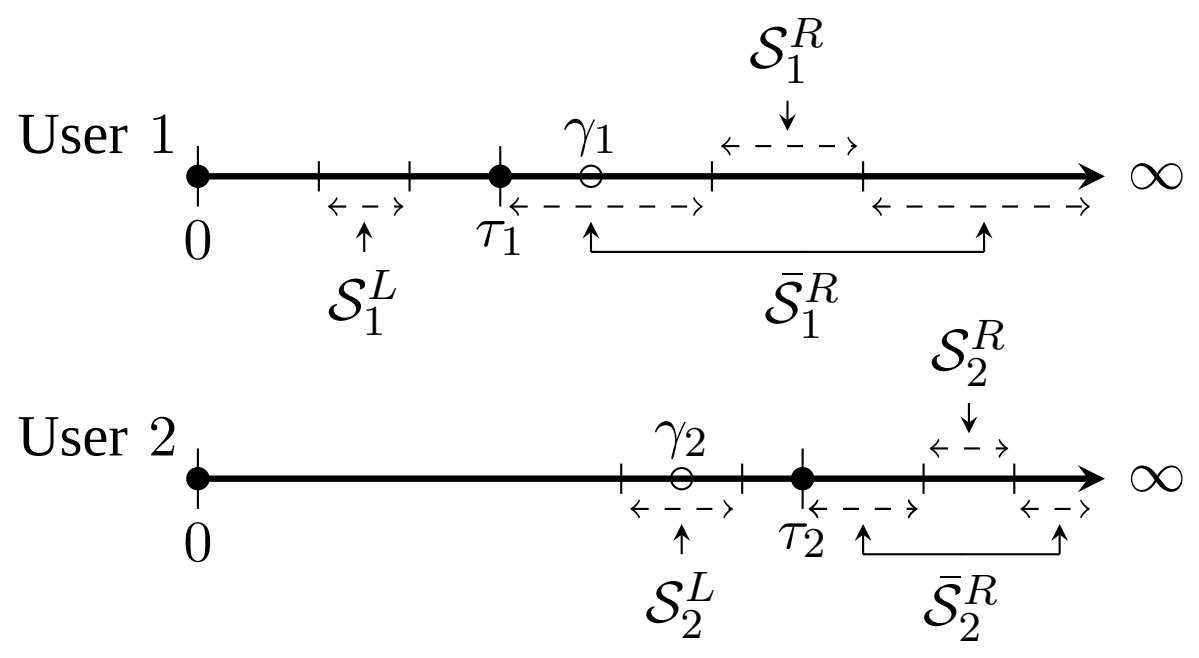

(a) $\gamma_1 \in \bar{\mathcal{S}}_1^R$, $\gamma_2 \in \mathcal{S}_2^L$ and $r(\boldsymbol{\mathcal{T}}, \boldsymbol{\Gamma}) < r(\boldsymbol{\mathcal{F}}, \boldsymbol{\Gamma})$.

(b) $\gamma_1 \in \bar{\mathcal{S}}_1^R$, $\gamma_2 \in \mathcal{S}_2^L$ and $r(\boldsymbol{\mathcal{T}}, \boldsymbol{\Gamma}) > r(\boldsymbol{\mathcal{F}}, \boldsymbol{\Gamma})$.

Fig. 1. A two-user example indicating problem complexity due to heterogeneity and the coupling effects between individual feedback policies.

To this end, we let

$$\mathcal{G}_1^{-1}(\boldsymbol{\mathcal{F}}(\boldsymbol{\Gamma})) = \{i \in \mathcal{N} : i \neq 1 \ \&\ i \in \mathcal{G}_1(\boldsymbol{\mathcal{F}}(\boldsymbol{\Gamma}))\}$$

for a given $\boldsymbol{\mathcal{F}} = (\mathcal{F}_1, \ldots, \mathcal{F}_n)^\top$. That is, $\mathcal{G}_1^{-1}(\boldsymbol{\mathcal{F}}(\boldsymbol{\Gamma}))$ is the random set of users containing all MUs requesting beam 1 under $\boldsymbol{\mathcal{F}}$, except for the first MU. The superscript $-1$ is used to indicate that all MUs but MU 1 requesting beam 1 are included in $\mathcal{G}_1^{-1}(\boldsymbol{\mathcal{F}}(\boldsymbol{\Gamma}))$. The maximum beam 1 SINR value achieved by a MU in this random set is denoted by $\bar{\gamma}_1^\star(\boldsymbol{\mathcal{F}})$, *i.e.*,$\bar{\gamma}_1^\star(\boldsymbol{\mathcal{F}}) = \max_{i \in \mathcal{G}_1^{-1}(\boldsymbol{\mathcal{F}}(\boldsymbol{\Gamma}))} \gamma_{1,i}$.

Consider now the decentralized feedback policy $\boldsymbol{\mathcal{F}}^1 = (\mathcal{T}_1, \mathcal{F}_2, \cdots, \mathcal{F}_n)^\top$. That is, we only allow MU 1 to switch to the threshold feedback rule $\mathcal{T}_1$ with the threshold value $\tau_1$ determined as above. Then, for almost all realizations of $\boldsymbol{\Gamma}$, we have $\bar{\gamma}_1^\star(\boldsymbol{\mathcal{F}}) = \bar{\gamma}_1^\star(\boldsymbol{\mathcal{F}}^1) = \bar{\gamma}_1^\star$. Therefore, the difference between $R(\boldsymbol{\mathcal{F}})$ and $R(\boldsymbol{\mathcal{F}}^1)$ depends only on the rate achieved by MU 1 under these two feedback policies.

We are interested in proving $R(\boldsymbol{\mathcal{F}}) \leq R(\boldsymbol{\mathcal{T}})$. A brief sketch of the proof is as follows. We first prove that $R(\boldsymbol{\mathcal{F}}) \leq R(\boldsymbol{\mathcal{F}}^1)$. To this end, we let $\boldsymbol{\Gamma}_{-1}$ be the SINR matrix containing SINR values of all MUs except those of the first MU. We also let $R(\boldsymbol{\mathcal{F}}|\boldsymbol{\Gamma}_{-1}) = \mathsf{E}_{\boldsymbol{\Gamma}}[r(\boldsymbol{\mathcal{F}}, \boldsymbol{\Gamma})|\boldsymbol{\Gamma}_{-1}]$ be the conditional average sum-rate achieved by $\boldsymbol{\mathcal{F}}$ for a given $\boldsymbol{\Gamma}_{-1}$. Then, it is enough to show that $R(\boldsymbol{\mathcal{F}}^1|\boldsymbol{\Gamma}_{-1}) \geq R(\boldsymbol{\mathcal{F}}|\boldsymbol{\Gamma}_{-1})$ for almost all $\boldsymbol{\Gamma}_{-1}$. This result implies that the sum-rate increases if MU 1 switches to a threshold feedback rule regardless of feedback rules of other MUs. Repeating the same steps for other MUs $i \in \{2, 3, \cdots, n\}$ one-by-one, we end up with the threshold feedback policy $\boldsymbol{\mathcal{T}}$ after $n$ steps, and conclude that $R(\boldsymbol{\mathcal{T}}) \geq R(\boldsymbol{\mathcal{F}})$.

Before giving the details of the proof sketched above, we will first perform a preliminary analysis. For the rest of this part of the paper, $\boldsymbol{\mathcal{F}}^1$ will represent the decentralized feedback policy derived from a given decentralized feedback policy $\boldsymbol{\mathcal{F}}$ as above. When we switch from $\boldsymbol{\mathcal{F}}$ to $\boldsymbol{\mathcal{F}}^1$, we can identify three main types of events: *neutral*, *loss* and *gain* events. On the neutral event, we will continue to achieve the same downlink throughput under both feedback policies. On the loss event, we will lose some data rate upon switching to $\boldsymbol{\mathcal{F}}^1$ from $\boldsymbol{\mathcal{F}}$. Finally, on the gain event, we will gain some data rate upon switching to $\boldsymbol{\mathcal{F}}^1$ from $\boldsymbol{\mathcal{F}}$. The difference $R\left(\boldsymbol{\mathcal{F}}^1\right) - R\left(\boldsymbol{\mathcal{F}}\right)$ depends on the average sum-rates lost and gained on the loss and gain events. To show that $R\left(\boldsymbol{\mathcal{F}}^1\right) - R\left(\boldsymbol{\mathcal{F}}\right) \geq 0$, we need to characterize these loss and gain events precisely. We first formally define these events, and then provide their further characterizations suitable for our analysis in Lemmas 1 and 2.

*Definition 5:* The loss, gain and neutral events upon switching to $\boldsymbol{\mathcal{F}}^1$ from $\boldsymbol{\mathcal{F}}$ on beam $1$ are defined as

$$A_L = \left\{ \boldsymbol{\Gamma} \in \mathbb{R}_+^{M \times n} \ : \ r^1\left(\boldsymbol{\mathcal{F}}^1, \boldsymbol{\Gamma}\right) < r^1\left(\boldsymbol{\mathcal{F}}, \boldsymbol{\Gamma}\right) \right\}, \tag{7}$$

$$A_G = \left\{ \boldsymbol{\Gamma} \in \mathbb{R}_+^{M \times n} \ : \ r^1\left(\boldsymbol{\mathcal{F}}^1, \boldsymbol{\Gamma}\right) > r^1\left(\boldsymbol{\mathcal{F}}, \boldsymbol{\Gamma}\right) \right\} \tag{8}$$

and

$$A_N = \left\{ \boldsymbol{\Gamma} \in \mathbb{R}_+^{M \times n} \ : \ r^1(\boldsymbol{\mathcal{F}}^1, \boldsymbol{\Gamma}) = r^1(\boldsymbol{\mathcal{F}}, \boldsymbol{\Gamma}) \right\}, \tag{9}$$

respectively.

The neutral event is not so much of an interest since both policies achieve the same rate on this event. However, loss and gain events require further evaluation, and the next two lemmas provide other characterizations for these events. These characterizations will be important when we compare $R\left(\boldsymbol{\mathcal{F}}^1\right)$ against $R\left(\boldsymbol{\mathcal{F}}\right)$.

*Lemma 1:* $A_L$ is equal to

$$A_L = \left\{ \boldsymbol{\Gamma} \in \mathbb{R}_+^{M \times n} \ : \ \boldsymbol{\gamma}_1 \in \mathcal{S}_1^L \quad \& \quad \bar{\gamma}_1^\star < \gamma_{1,1} \right\}.$$

*Proof:* See Appendix A. ∎

A similar characterization for the gain event on beam $1$ is given in the next lemma.

*Lemma 2:* $A_G$ is equal to

$$A_G = \left\{ \boldsymbol{\Gamma} \in \mathbb{R}_+^{M \times n} \ : \ \boldsymbol{\gamma}_1 \in \bar{\mathcal{S}}_1^R \quad \& \quad \bar{\gamma}_1^\star < \gamma_{1,1} \right\}.$$

*Proof:* See Appendix A. ∎

These auxiliary results will aid to prove sum-rate optimality of $\boldsymbol{\mathcal{F}}^1$ over $\boldsymbol{\mathcal{F}}$ in Theorem 1. Before providing the details of the proof of this theorem, we will again give a sketch of the proof. $A_L$, $A_G$ and $A_N$ are three disjoint events with total probability mass of one. Therefore, for a feedback policy $\boldsymbol{\mathcal{F}}$, we can write $R^1(\boldsymbol{\mathcal{F}}|\boldsymbol{\Gamma}_{-1}) = R^1(\boldsymbol{\mathcal{F}}, A_L|\boldsymbol{\Gamma}_{-1}) + R^1(\boldsymbol{\mathcal{F}}, A_G|\boldsymbol{\Gamma}_{-1}) + R^1(\boldsymbol{\mathcal{F}}, A_N|\boldsymbol{\Gamma}_{-1})$.

We can write a similar expression for $R^1\left(\boldsymbol{\mathcal{F}}^1|\boldsymbol{\Gamma}_{-1}\right)$. Comparison of these two expressions term-by-term reveals that $R^1\left(\boldsymbol{\mathcal{F}}^1|\boldsymbol{\Gamma}_{-1}\right) \geq R^1(\boldsymbol{\mathcal{F}}|\boldsymbol{\Gamma}_{-1})$. Since this inequality holds for almost all $\boldsymbol{\Gamma}_{-1}$, we also have $R^1\left(\boldsymbol{\mathcal{F}}^1\right) \geq R^1(\boldsymbol{\mathcal{F}})$. Since beams are statistically identical, the total rate is $M$ times the rate achieved on beam 1. Therefore, we finally have $R\left(\boldsymbol{\mathcal{F}}^1\right) \geq R(\boldsymbol{\mathcal{F}})$. We make this idea formal in the proof of the next theorem.

*Theorem 1:* Let $\boldsymbol{\mathcal{F}} = (\mathcal{F}_1, \ldots, \mathcal{F}_n)^\top$ and $\boldsymbol{\mathcal{F}}^1 = (\mathcal{T}_1, \mathcal{F}_2, \ldots, \mathcal{F}_n)^\top$ be defined as above. Then, $\Lambda(\boldsymbol{\mathcal{F}}) = \Lambda\left(\boldsymbol{\mathcal{F}}^1\right)$, and $R\left(\boldsymbol{\mathcal{F}}^1\right) \geq R(\boldsymbol{\mathcal{F}})$ for any $M \geq 1$.

*Proof:* It is enough to prove $R^1\left(\boldsymbol{\mathcal{F}}^1|\boldsymbol{\Gamma}_{-1}\right) \geq R^1(\boldsymbol{\mathcal{F}}|\boldsymbol{\Gamma}_{-1})$ for almost all $\boldsymbol{\Gamma}_{-1}$. By definition, we have $R^1(\boldsymbol{\mathcal{F}}, A_N|\boldsymbol{\Gamma}_{-1}) = R^1\left(\boldsymbol{\mathcal{F}}^1, A_N|\boldsymbol{\Gamma}_{-1}\right)$, and therefore we are only interested in the average sum-rates on loss and gain events.

The following identity follows from the definition of conditional expectation.

$$R^1(\boldsymbol{\mathcal{F}}, A_L|\boldsymbol{\Gamma}_{-1}) = \mathsf{Pr}(A_L|\boldsymbol{\Gamma}_{-1})\, \mathsf{E}_{\boldsymbol{\Gamma}}\left[r^1(\boldsymbol{\mathcal{F}}, \boldsymbol{\Gamma})\,|A_L, \boldsymbol{\Gamma}_{-1}\right].$$

Lemma 1 implies that whenever $A_L$ is correct, MU 1 requests beam 1, and achieves the best SINR on beam 1 among all the MUs requesting beam 1. Since $\boldsymbol{\gamma}_1 \in \mathcal{S}_1^L$ on $A_L$, $\gamma_{1,1}$ is less than $\tau_1$. Therefore,

$$R^1(\boldsymbol{\mathcal{F}}, A_L|\boldsymbol{\Gamma}_{-1}) \quad \leq \quad \mathsf{Pr}(A_L|\boldsymbol{\Gamma}_{-1}) \log(1+\tau_1). \tag{10}$$

Similarly, we can write

$$R^1(\boldsymbol{\mathcal{F}}, A_G|\boldsymbol{\Gamma}_{-1}) = \mathsf{Pr}(A_G|\boldsymbol{\Gamma}_{-1})\, \mathsf{E}_{\boldsymbol{\Gamma}}\left[r^1(\boldsymbol{\mathcal{F}}, \boldsymbol{\Gamma})\,|A_G, \boldsymbol{\Gamma}_{-1}\right].$$

Lemma 2 implies that MU 1 achieves the best SINR on beam 1 among all the MUs requesting beam 1 but $\boldsymbol{\gamma}_1 \in \bar{\mathcal{S}}_1^R$ on $A_G$. Therefore, $\boldsymbol{\gamma}_1 \notin FB_1$, and MU 1 will not request beam 1 under $\boldsymbol{\mathcal{F}}$. Hence, $\boldsymbol{\mathcal{F}}$ schedules beam 1 to the MU with SINR value $\bar{\gamma}_1^\star$, which leads to [3]

$$R^1(\boldsymbol{\mathcal{F}}, A_G|\boldsymbol{\Gamma}_{-1}) \quad = \quad \mathsf{Pr}(A_G|\boldsymbol{\Gamma}_{-1}) \log(1+\bar{\gamma}_1^\star). \tag{11}$$

Similar to the above arguments, MU 1 will not request beam 1 under $\boldsymbol{\mathcal{F}}^1$ on the event $A_L$ since $\boldsymbol{\gamma}_1 \in \mathcal{S}_1^L$. This means

$$R^1\left(\boldsymbol{\mathcal{F}}^1, A_L|\boldsymbol{\Gamma}_{-1}\right) = \mathsf{Pr}(A_L|\boldsymbol{\Gamma}_{-1}) \log(1+\bar{\gamma}_1^\star). \tag{12}$$

[3]Note that $\bar{\gamma}_1^\star$ is a (measurable) function of $\boldsymbol{\Gamma}_{-1}$, and therefore (11) conforms with the measure theoretic definition of the conditional expectation.

Finally, MU 1 requests beam 1 under $\boldsymbol{\mathcal{F}}^1$ on $A_G$, leading to

$$R^1\left(\boldsymbol{\mathcal{F}}^1, A_G|\boldsymbol{\Gamma}_{-1}\right) \geq \mathsf{Pr}\left(A_G|\boldsymbol{\Gamma}_{-1}\right)\log\left(1+\max\left(\tau_1, \bar{\gamma}_1^\star\right)\right). \quad (13)$$

By using (10), (11), (12) and (13), we have

$$R^1\left(\boldsymbol{\mathcal{F}}^1|\boldsymbol{\Gamma}_{-1}\right) - R^1\left(\boldsymbol{\mathcal{F}}|\boldsymbol{\Gamma}_{-1}\right) \geq \mathsf{Pr}\left(A_G|\boldsymbol{\Gamma}_{-1}\right)\left(\log\left(1+\max\left(\tau_1,\bar{\gamma}_1^\star\right)\right) - \log\left(1+\bar{\gamma}_1^\star\right)\right) + \mathsf{Pr}\left(A_L|\boldsymbol{\Gamma}_{-1}\right)\left(\log\left(1+\bar{\gamma}_1^\star\right) - \log\left(1+\tau_1\right)\right).$$

To conclude the proof, we need to analyze two different cases separately. If $\bar{\gamma}_1^\star \geq \tau_1$, then it directly follows that $R^1\left(\boldsymbol{\mathcal{F}}^1|\boldsymbol{\Gamma}_{-1}\right) - R^1\left(\boldsymbol{\mathcal{F}}|\boldsymbol{\Gamma}_{-1}\right) \geq 0$. If $\bar{\gamma}_1^\star < \tau_1$, then we have

$$R^1\left(\boldsymbol{\mathcal{F}}^1|\boldsymbol{\Gamma}_{-1}, \bar{\gamma}_1^\star < \tau_1\right) - R^1\left(\boldsymbol{\mathcal{F}}|\boldsymbol{\Gamma}_{-1}, \bar{\gamma}_1^\star < \tau_1\right) \geq \left(\mathsf{Pr}\left(A_G|\boldsymbol{\Gamma}_{-1}, \bar{\gamma}_1^\star < \tau_1\right) - \mathsf{Pr}\left(A_L|\boldsymbol{\Gamma}_{-1}, \bar{\gamma}_1^\star < \tau_1\right)\right)\left(\log\left(1+\tau_1\right) - \log\left(1+\bar{\gamma}_1^\star\right)\right).$$

Observe that $\mathsf{Pr}\left(A_G|\boldsymbol{\Gamma}_{-1}, \bar{\gamma}_1^\star < \tau_1\right) = \mathsf{Pr}\left\{\boldsymbol{\gamma}_1 \in \bar{\mathcal{S}}_1^R\right\}$ and $\mathsf{Pr}\left\{A_L|\boldsymbol{\Gamma}_{-1}, \bar{\gamma}_1^\star < \tau_1\right\} \leq \mathsf{Pr}\left\{\boldsymbol{\gamma}_1 \in \mathcal{S}_1^L\right\}$. Since $\mathsf{Pr}\left\{\boldsymbol{\gamma}_1 \in \bar{\mathcal{S}}_1^R\right\} = \mathsf{Pr}\left\{\boldsymbol{\gamma}_1 \in \mathcal{S}_1^L\right\}$, we have $R^1\left(\boldsymbol{\mathcal{F}}^1|\boldsymbol{\Gamma}_{-1}, \bar{\gamma}_1^\star < \tau_1\right) - R^1\left(\boldsymbol{\mathcal{F}}|\boldsymbol{\Gamma}_{-1}, \bar{\gamma}_1^\star < \tau_1\right) \geq 0$. After removing conditioning, this proves that $R^1\left(\boldsymbol{\mathcal{F}}^1|\boldsymbol{\Gamma}_{-1}\right) \geq R^1\left(\boldsymbol{\mathcal{F}}|\boldsymbol{\Gamma}_{-1}\right)$ for almost all $\boldsymbol{\Gamma}_{-1}$, and therefore $R^1\left(\boldsymbol{\mathcal{F}}^1\right) \geq R^1\left(\boldsymbol{\mathcal{F}}\right)$. ∎

This theorem shows that if a MU starts using a threshold feedback rule, the sum-rate improves regardless of the feedback rules of all other users. This leads to the following key finding.

*Theorem 2:* For any beam symmetric general decentralized feedback policy $\boldsymbol{\mathcal{F}}$, there exists a GTFP $\boldsymbol{\mathcal{T}}$ such that $\Lambda\left(\boldsymbol{\mathcal{F}}\right) = \Lambda\left(\boldsymbol{\mathcal{T}}\right)$ and $R\left(\boldsymbol{\mathcal{T}}\right) \geq R\left(\boldsymbol{\mathcal{F}}\right)$.

*Proof:* For a given $\boldsymbol{\mathcal{F}} = \left(\mathcal{F}_1, \ldots, \mathcal{F}_n\right)^\top$, let $\boldsymbol{\mathcal{T}} = \left(\mathcal{T}_1, \ldots, \mathcal{T}_n\right)^\top$ be the GTFP constructed as above. Let $\boldsymbol{\mathcal{F}}^k = \left(\mathcal{T}_1, \ldots, \mathcal{T}_k, \mathcal{F}_{k+1}, \ldots, \mathcal{F}_n\right)^\top$ for $1 \leq k \leq n-1$. When $k = n$, we have $\boldsymbol{\mathcal{F}}^n = \boldsymbol{\mathcal{T}}$. By Theorem 1, we have $R\left(\boldsymbol{\mathcal{F}}\right) \leq R\left(\boldsymbol{\mathcal{F}}^1\right) \leq \cdots \leq R\left(\boldsymbol{\mathcal{F}}^n\right) = R\left(\boldsymbol{\mathcal{T}}\right)$. Since $\Lambda\left(\boldsymbol{\mathcal{F}}\right) = \Lambda\left(\boldsymbol{\mathcal{F}}^1\right) = \cdots = \Lambda\left(\boldsymbol{\mathcal{F}}^n\right) = \Lambda\left(\boldsymbol{\mathcal{T}}\right)$, the proof is complete. ∎

### *B. Optimality of Maximum* SINR *Threshold Feedback Policies*

In this part, we briefly explain why similar results also hold for MTFPs. The proof techniques are the same except for some subtle differences. To start with, under a maximum SINR decentralized feedback policy, each MU requests only the beam achieving the maximum SINR if the feedback conditions are met, *i.e.*, see Definitions 1 and 4. Hence, the thresholds are set such that $\mathsf{Pr}\left\{b_i^\star = 1 \text{ and } \gamma_i^\star \geq \tau_i\right\} = \mathsf{Pr}\left\{\boldsymbol{\gamma_i} \in FB_i\right\}$. The definition of $FB_i$ is refined in which MU $i$ requests beam 1 if and only if $b_i^\star = 1$ and $\gamma_i^\star$ satisfies feedback conditions. The definitions of other sets and events of interest require only some subtle modifications, too. For example, $A_L$ can now be defined as

$$A_L = \left\{ \mathbf{\Gamma} \in \mathbb{R}_+^{M \times n} \ : \ \boldsymbol{\gamma}_1 \in \mathcal{S}_1^L \quad \& \quad \bar{\gamma}_1^\star < \gamma_1^\star \right\},$$

where $\mathcal{S}_1^L = \left\{ \boldsymbol{\gamma}_1 \in \mathbb{R}_+^M : \boldsymbol{\gamma}_1 \in FB_1 \ \& \ \gamma_1^\star < \tau_1 \right\}$. The next two theorems provide results analogous to the ones stated in Theorems 1 and 2.

*Theorem 3:* For a given beam symmetric decentralized maximum SINR policy $\boldsymbol{\mathcal{F}} = (\mathcal{F}_1, \ldots, \mathcal{F}_n)^\top$, let $\boldsymbol{\mathcal{F}}^1 = (\mathcal{T}_1, \mathcal{F}_2, \ldots, \mathcal{F}_n)^\top$ be the maximum SINR threshold feedback policy derived from $\boldsymbol{\mathcal{F}}$ by allowing MU 1 to switch from $\mathcal{F}_1$ to $\mathcal{T}_1$, where $\mathcal{T}_1$ is a beam symmetric maximum SINR threshold rule whose threshold is set as above. Then, $\Lambda(\boldsymbol{\mathcal{F}}) = \Lambda\left(\boldsymbol{\mathcal{F}}^1\right)$, and $R\left(\boldsymbol{\mathcal{F}}^1\right) \geq R(\boldsymbol{\mathcal{F}})$ for any $M \geq 1$.

*Theorem 4:* For any beam symmetric decentralized maximum SINR feedback policy $\boldsymbol{\mathcal{F}}$, there exists an MTFP $\boldsymbol{\mathcal{T}}$ such that $\Lambda(\boldsymbol{\mathcal{F}}) = \Lambda(\boldsymbol{\mathcal{T}})$ and $R(\boldsymbol{\mathcal{T}}) \geq R(\boldsymbol{\mathcal{F}})$.

Since the proofs of these theorems are similar to the proofs above, we skip them to avoid repetition. It is important to note that Theorems 2 and 4 hold for any continuous SINR distribution.

### C. Discussion of Results

In this part, we briefly discuss the results presented above. We start with a comparison between GTFPs and MTFPs. The main advantage of GTFPs over MTFPs is the ability of the BS to allocate multiple beams to a MU. Therefore, a GTFP policy achieves higher data rates when compared to an MTFP policy with the same threshold levels. From a practical point of view, such gains in data rates are expected to be minor due to dependencies among beams at a MU, *i.e.,* high $\gamma_{m,i}$ implies low $\gamma_{k,i}$, $\forall k \neq m$. Moreover, both types of policies achieve the same performance if all threshold values are greater than 1, which is formally proved in the next lemma.

*Lemma 3:* Let $\boldsymbol{\mathcal{T}}$ be an MTFP with thresholds $\{\tau_i\}_{i \in \mathcal{N}}$, and $\boldsymbol{\mathcal{T}}'$ be the corresponding GTFP with the same threshold levels. Let $\mathcal{N}_m$ and $\mathcal{N}_m'$ be the sets of MUs requesting beam $m \in \mathcal{M}$ according to $\boldsymbol{\mathcal{T}}$ and $\boldsymbol{\mathcal{T}}'$, respectively. If $\tau_i > 1$ for all $i \in \mathcal{N}$, then $\mathcal{N}_m = \mathcal{N}_m'$.

*Proof:* See Appendix B. ■

Note that the requirement on threshold values for the equality of MTFPs and GTFPs in Lemma 3 is only a $0$ [dB] requirement, which is practically a quite low SINR value. This implies that both feedback policies will actually achieve the same sum-rate in almost all practical communication scenarios.

On the other hand, from a theoretical point of view, the resulting optimization problem over $\mathbb{R}_+^n$ lends itself more amenable to further analysis if we only focus on GTFPs. More specifically, we can search for the optimal beam symmetric feedback policies within the class of GTFPs without sacrificing from optimality thanks to Theorem 2, and with a slight abuse of notation, we can equivalently write (5) as

$$\begin{array}{ll} \underset{\boldsymbol{\tau} \in \mathbb{R}_+^n}{\text{maximize}} & R(\boldsymbol{\tau}) \\ \text{subject to} & \sum_{i=1}^{n} \mathsf{Pr}\{\gamma_{1,i} \geq \tau_i\} \leq \lambda \end{array}. \tag{14}$$

Some further game theoretic insights are as follows. We will only focus on GTFPs but similar explanations also hold for MTFPs. Given the same utility function $R(\mathcal{F}_1, \ldots, \mathcal{F}_n)$ for all MUs, the selfish optimization problem faced by MU $i$ is to choose a beam symmetric decentralized feedback rule maximizing its utility given other MUs' feedback rules without increasing the feedback level. Theorem 1 shows that the *dominant strategy* is to switch from $\mathcal{F}_i$ to the corresponding threshold rule $\mathcal{T}_i$. As a result, the set of GTFPs constitute the set of Nash equilibria for this feedback rule selection game, and therefore GTFPs are also stable operating points from a game theoretic point of view.

In the rest of the paper, we will analyze the finite dimensional optimization problem in (14). We will show that the sum-rate becomes a Schur-concave function of feedback probabilities $p_i = \mathsf{Pr}\{\gamma_{1,i} \geq \tau_i\}$ if the $\mathrm{SINR}$ distribution satisfies some mild conditions. This result establishes the optimality of homogenous general threshold feedback policies among the class of beam symmetric general decentralized feedback policies.

## V. Optimal Threshold Selection Problem

The optimization problem in (14), which we call *optimal threshold selection problem*, is still not easy to solve, even for a simple two-user system, due to the non-convex objective function and the non-convex constraint set depending on the distribution of $\mathrm{SINR}$ values. The complexity of the problem further increases with increasing numbers of users due to the dimensionality growth. Therefore, it is not possible to solve the optimal threshold selection problem in its full generality for a general $n$-user system. However, we can still search for a structure in the sum-rate function to solve the optimal threshold selection problem, which is what we will do in the remainder of this section.

More specifically, we will search for sufficient conditions to be satisfied by $\mathrm{SINR}$ distributions so that the sum-rate becomes a Schur-concave function of feedback probabilities. Roughly speaking, a Schur-concave function increases when the dispersion among the components of its argument decreases, which implies a solution for the optimization problem in (14) is a homogenous threshold feedback policy in which thresholds are set according to

$$\boldsymbol{\tau}^{\star} = \left(F^{-1}\left(1 - \frac{\lambda}{n}\right), \cdots, F^{-1}\left(1 - \frac{\lambda}{n}\right)\right)^{\top} \tag{15}$$

if the sum-rate is a Schur-concave function. We make this intuitive idea rigorous below.

The rest of this section is organized as follows. We first provide an overview of our main results in the next subsection without any formal proofs. We then introduce some key concepts from the theory of majorization in Subsection V-B. Finally, formal proofs are supplied in Subsection V-C and in related appendices.

## *A. Main Results*

The main results of this section are stated in Theorems 5 and 6. In these theorems, we view the sum-rate as a function of feedback probabilities. This approach does not limit the generality of our results since $\mathrm{SINR}$ probability density function is already assumed to have $\mathbb{R}_+$ as its support, and therefore there is a one-to-one correspondence between feedback threshold values $\tau_i$ and the feedback probabilities $p_i$, *i.e.*, $\tau_i = F^{-1}(1-p_i)$ for all $i \in \mathcal{N}$. As already noted in Section III, this assumption is satisfied for many commonly used practical fading models such as Rayleigh, Rician and Nakagami fading. It may still be possible to extend similar proof techniques to more general fading distributions; a future research direction of interest which we do not pursue in this paper since the analysis is already complicated even with this simplifying assumption. Our theorems are as follows.

*Theorem 5:* The sum-rate $R(\boldsymbol{p})$ is a Schur-concave function if

$$\log(1+\gamma)(\lambda-2q) + \int_{F^{-1}(1+q-\lambda)}^{F^{-1}(1-q)} \frac{F(x)}{1+x} dx - (\lambda-2q)\log\left(1+F^{-1}(1+q-\lambda)\right) \geq 0 \tag{16}$$

for all $\gamma \geq 0$, $\lambda \in [0,2]$ and $\max\{0, \lambda-1\} \leq q \leq \frac{\lambda}{2}$.

*Theorem 6:* The sum-rate $R(\boldsymbol{p})$ is Schur-concave if $f$ is bounded at zero, and has the derivative $f'$ satisfying

$$f'\left(F^{-1}(x)\right) \leq -\frac{f\left(F^{-1}(x)\right)}{1+F^{-1}(x)} \tag{17}$$

for all $x \in [0,1]$.

The proofs of Theorems 5 and 6 require introduction of new notation, and involve several cases to analyze separately. We also need some key results from the theory of majorization [19] to prove these results. Therefore, we have relegated their proofs to the following subsections and appendixes. We now briefly discuss their implications.

We first note that the sufficient condition for the Schur-concavity of the sum-rate given in (16) is stronger than the one given in (17) in the sense that (16) always holds whenever (17) holds, but not vice versa. This is formally established in Subsection V-C. Furthermore, since the first term in (16) is always positive, an easier condition to check for the Schur-concavity of the sum-rate function is

$$\int_{F^{-1}(1+q-\lambda)}^{F^{-1}(1-q)} \frac{F(x)}{1+x}dx - (\lambda - 2q)\log\left(1 + F^{-1}(1+q-\lambda)\right) \geq 0 \tag{18}$$

for all $\lambda \in [0,2]$ and $\max\{0, \lambda - 1\} \leq q \leq \frac{\lambda}{2}$. Further, we can bound (18) from below to obtain another sufficient condition as

$$(1+q-\lambda)\log\left(1+F^{-1}(1-q)\right) - (1-q)\log\left(1+F^{-1}\left(1+q-\lambda\right)\right) \geq 0, \tag{19}$$

for all $\lambda \in [0,2]$ and $\max\{0, \lambda - 1\} \leq q \leq \frac{\lambda}{2}$. For a two-user system, (18) is also necessary, *i.e.*, see Lemma 10 and discussions therein.

Although the conditions (18) and (19) are easy to verify numerically, they may not be tractable analytically. The integral expression in (18) is hard to evaluate in closed-form. Analytical verification of (19) is also difficult due to the presence of conflicting forces working in opposite directions to increase/decrease the value of the bound. For example, the pre-log factor of the first term in (19), which is $1+q-\lambda$, is smaller than the pre-log factor of the second term, which is $1-q$, for $\max\{0, \lambda - 1\} \leq q \leq \frac{\lambda}{2}$. Conversely, for $\max\{0, \lambda - 1\} \leq q \leq \frac{\lambda}{2}$, $F^{-1}(1-q)$ appearing inside the logarithm in the first term is greater than $F^{-1}(1+q-\lambda)$ appearing inside the logarithm in the second term.

On the other hand, the sufficient condition for the Schur-concavity of the sum-rate function given in Theorem 6 turns out to be much easier to deal with analytically although it looks more complex than (18) and (19). In particular, it provides an *almost* complete characterization for the solution of optimal threshold selection problem for richly scattered Rayleigh fading environments. More precisely, (17) is always satisfied for all values of $\rho$ for Rayleigh fading channels whenever $M \geq 2$. Hence, the sum-rate is always a Schur-concave function of feedback probabilities in this case, and is maximized if thresholds are chosen according to (15). In Section VI, we provide a detailed discussion for the optimality and sub-optimality of homogenous threshold feedback policies for Rayleigh fading channels as well as other wireless channel models. Next, we will briefly introduce some key concepts from the theory of majorization to be used later in our analysis.

### B. Majorization

For a vector $\boldsymbol{x}$ in $\mathbb{R}^n$, we denote its ordered coordinates by $x_{(1)} \geq \cdots \geq x_{(n)}$. For $\boldsymbol{x}$ and $\boldsymbol{y}$ in $\mathbb{R}^n$, we say $\boldsymbol{x}$ *majorizes* $\boldsymbol{y}$ and write it as $\boldsymbol{x} \succeq_{\mathrm{M}} \boldsymbol{y}$ if we have $\sum_{i=1}^{k} x_{(i)} \geq \sum_{i=1}^{k} y_{(i)}$ when $k = 1, \ldots, n-1$, and $\sum_{i=1}^{n} x_{(i)} = \sum_{i=1}^{n} y_{(i)}$. A function $\varphi : \mathbb{R}^n \mapsto \mathbb{R}$ is said to be *Schur-convex* if $\boldsymbol{x} \succeq_{\mathrm{M}} \boldsymbol{y}$ implies $\varphi(\boldsymbol{x}) \geq \varphi(\boldsymbol{y})$, and $\varphi$ is *Schur-concave* if $-\varphi$ is Schur-convex. Schur-convex/concave functions often arise in mathematical analysis and engineering applications [30], [31]. For example, every function that is concave (convex) and symmetric is also a Schur-concave (Schur-convex) function.

A Schur-concave function tends to increase when the components of its argument become more similar. We will establish conditions under which the sum-rate becomes a Schur-concave function, which will, in turn, imply the optimality of homogenous threshold feedback policies. The following lemma is helpful in establishing these conditions.

*Lemma 4:* Let $\varphi$ be a real-valued function defined on $\mathbb{R}_+^n$, and $\mathcal{D} = \left\{\boldsymbol{z} \in \mathbb{R}_+^n : z_1 \geq \cdots \geq z_n\right\}$. Then, $\varphi$ is a Schur-convex function if and only if, for all $\boldsymbol{z} \in \mathcal{D}$ and $i = 1, \ldots, n-1$,

$$\varphi\left(z_1, \ldots, z_{i-1}, z_i + \epsilon, z_{i+1} - \epsilon, z_{i+2}, \ldots, z_n\right)$$

is increasing in $\epsilon$ over the region $0 \leq \epsilon \leq \min\left\{z_{i-1} - z_i, z_{i+1} - z_{i+2}\right\}$. [4]

It can be seen that the coordinates $z_i$ and $z_{i+1}$ are systematically altered by using the parameter $\epsilon$, and the constraints on $\epsilon$ eliminate any violation in the order. Interested readers are referred to [19] for more insights on the theory of majorization. Now, we will see how we can use this theory to identify the Schur-concave structure in the objective rate function.

### C. Schur-concavity Analysis for the Sum-rate

The main objective is to establish sufficient conditions on the $\mathrm{SINR}$ distributions for the Schur-concavity of the sum-rate function. Again, we focus on the first beam to explain our proof ideas without any loss of generality since all beams are statistically identical. We start by analyzing the sum-rate as a function of thresholds as given in (14) to establish three important lemmas. Next, we will incorporate the feedback constraint into our optimization problem by interpreting the sum-rate as a function of feedback probabilities. Using these results, we will finally establish the underlying Schur-concave structure in the sum-rate function through the theory of majorization.

*1) Rate as a Function of Thresholds:* Consider thresholds in increasing order, *i.e.*, $\tau_{\pi(1)} \leq \cdots \leq \tau_{\pi(i)} \leq \tau_{\pi(i+1)} \leq \cdots \leq \tau_{\pi(n)}$. Based on Lemma 4, it is enough to consider

$$R^1\left(\tau_{\pi(i+1)} + \epsilon, \tau_{\pi(i)} - \epsilon\right) = R^1\left(\tau_{\pi(n)}, \ldots, \tau_{\pi(i+1)} + \epsilon, \tau_{\pi(i)} - \epsilon, \ldots, \tau_{\pi(1)}\right),$$

[4]At the end points $i = 1$, $i = n - 1$, the condition is modified accordingly.

to identify the underlying Schur-concave structure in the sum-rate function.[5] However, analysis of this function is still complex. Therefore, we resort to the following divide-and-conquer approach.

Let $\mathcal{N}' = \{k \in \mathcal{N} : k \neq \pi(i) \ \& \ \pi(i+1)\}$. We fix the thresholds and the SINR values of all MUs in $\mathcal{N}'$.[6] Randomness is now associated only with MUs $\pi(i)$ and $\pi(i+1)$. With a slight abuse of notation, we define the truncated SINR on beam $m$ at MU $i$ as $\bar{\gamma}_{m,i} = \gamma_{m,i} 1_{\{\gamma_{m,i} \geq \tau_i\}}$. Let $\bar{\gamma}^{\star}_{\mathcal{N}'} = \max_{k \in \mathcal{N}'} \bar{\gamma}_{1,k}$, which is the maximum truncated SINR on beam 1 among the MUs in $\mathcal{N}'$. The instantaneous rate on beam 1 as a function of $\bar{\gamma}_{1,\pi(i)}$, $\bar{\gamma}_{1,\pi(i+1)}$ and $\bar{\gamma}^{\star}_{\mathcal{N}'}$ is

$$r^1\left(\bar{\gamma}_{1,\pi(i+1)}, \bar{\gamma}_{1,\pi(i)}, \bar{\gamma}^{\star}_{\mathcal{N}'}\right) = \log\left(1 + \max\left\{\bar{\gamma}_{1,\pi(i+1)}, \bar{\gamma}_{1,\pi(i)}, \bar{\gamma}^{\star}_{\mathcal{N}'}\right\}\right). \tag{20}$$

Therefore,

$$R^1\left(\tau_{\pi(i+1)}, \tau_{\pi(i)} | \bar{\gamma}^{\star}_{\mathcal{N}'}\right) = \mathsf{E}\left[r^1\left(\bar{\gamma}_{1,\pi(i+1)}, \bar{\gamma}_{1,\pi(i)}, \bar{\gamma}^{\star}_{\mathcal{N}'}\right) | \bar{\gamma}^{\star}_{\mathcal{N}'}\right]. \tag{21}$$

As shown later in the paper, this approach helps us to use the results derived for a two-user system to simplify our analysis. Therefore, considering a two-user system first, the rate on beam 1 as a function of the thresholds is explicitly given in Lemma 5.

*Lemma 5:* The rate on beam 1 of a two-user system is equal to

$$R^1(\boldsymbol{\tau}) = \int_{\tau_{\pi(2)}}^{\infty} \log(1+x) dF^2(x) + F\left(\tau_{\pi(2)}\right) \int_{\tau_{\pi(1)}}^{\tau_{\pi(2)}} \log(1+x) dF(x).$$

*Proof:* See Appendix C. ■

Coming back to the general $n$-user scenario, it is not tractable to obtain the rate explicitly as we have done in the previous lemma. However, we can explicitly write down an expression for $R^1\left(\tau_{\pi(i+1)}, \tau_{\pi(i)} | \bar{\gamma}^{\star}_{\mathcal{N}'}\right)$. $R^1\left(\tau_{\pi(i+1)}, \tau_{\pi(i)} | \bar{\gamma}^{\star}_{\mathcal{N}'}\right)$ is parameterized by $\bar{\gamma}^{\star}_{\mathcal{N}'}$, and its shape depends on the value of $\bar{\gamma}^{\star}_{\mathcal{N}'}$. Three cases of interest are $\bar{\gamma}^{\star}_{\mathcal{N}'} > \tau_{\pi(i+1)}$, $\bar{\gamma}^{\star}_{\mathcal{N}'} < \tau_{\pi(i)}$ and $\tau_{\pi(i)} \leq \bar{\gamma}^{\star}_{\mathcal{N}'} \leq \tau_{\pi(i+1)}$. We will now establish three important lemmas for these three cases, which will be useful in interpreting the rate function. The two-user rate expression given in Lemma 5 functions as a building block to obtain beam 1 rate expressions in these cases. We will start with the case $\bar{\gamma}^{\star}_{\mathcal{N}'} > \tau_{\pi(i+1)}$.

*Lemma 6:* If $\bar{\gamma}^{\star}_{\mathcal{N}'} > \tau_{\pi(i+1)}$, $R^1\left(\tau_{\pi(i+1)}, \tau_{\pi(i)} | \bar{\gamma}^{\star}_{\mathcal{N}'}\right)$ is given

$$R_0\left(\bar{\gamma}^{\star}_{\mathcal{N}'}\right) = \mathsf{Pr}\left\{\xi^{\star}_{i+1,i} \leq \bar{\gamma}^{\star}_{\mathcal{N}'} | \bar{\gamma}^{\star}_{\mathcal{N}'}\right\} \log\left(1 + \bar{\gamma}^{\star}_{\mathcal{N}'}\right) + \mathsf{E}\left[\log\left(1 + \xi^{\star}_{i+1,i}\right) 1_{\left\{\xi^{\star}_{i+1,i} > \bar{\gamma}^{\star}_{\mathcal{N}'}\right\}} | \bar{\gamma}^{\star}_{\mathcal{N}'}\right],$$

[5]We suppress the dependency of $R^1$ on $\tau_{\pi(k)}, k \neq i, i+1$ here and later in the paper when we focus only on thresholds $\tau_{\pi(i)}$ and $\tau_{\pi(i+1)}$.

[6]Fixing random SINR values means conditioning on them in the probabilistic sense. Indeed, it is sufficient just to condition on the maximum truncated beam 1 SINR value corresponding to MUs in $\mathcal{N}'$.

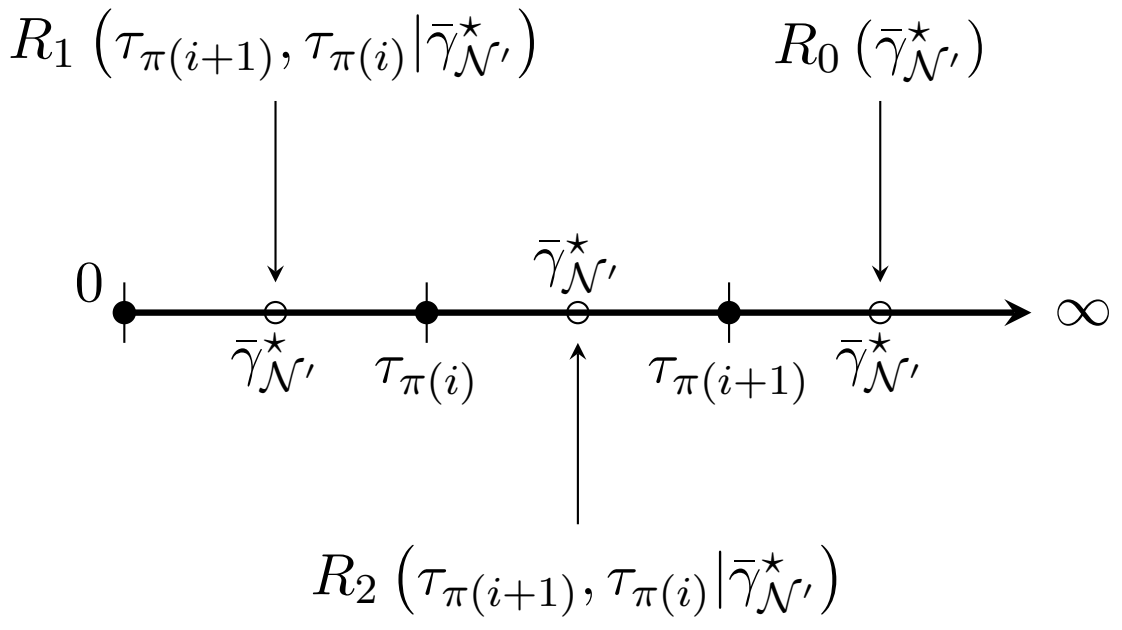

Fig. 2. Beam 1 rate as a function of thresholds for different values of $\bar{\gamma}^\star_{\mathcal{N}'}$

where $\xi^\star_{i+1,i} = \max\left\{\gamma_{1,\pi(i+1)}, \gamma_{1,\pi(i)}\right\}$.

*Proof:* See Appendix D. ■

Note that $R^1\left(\tau_{\pi(i+1)}, \tau_{\pi(i)}|\bar{\gamma}^\star_{\mathcal{N}'}\right)$ depends only on $\bar{\gamma}^\star_{\mathcal{N}'}$ but not on $\tau_{\pi(i)}$ and $\tau_{\pi(i+1)}$ when $\bar{\gamma}^\star_{\mathcal{N}'} > \tau_{\pi(i+1)}$. The next lemma provides an analogous expression for $R^1\left(\tau_{\pi(i+1)}, \tau_{\pi(i)}|\bar{\gamma}^\star_{\mathcal{N}'}\right)$ when $\bar{\gamma}^\star_{\mathcal{N}'} < \tau_{\pi(i)}$.

*Lemma 7:* If $\bar{\gamma}^\star_{\mathcal{N}'} < \tau_{\pi(i)}$, $R^1\left(\tau_{\pi(i+1)}, \tau_{\pi(i)}|\bar{\gamma}^\star_{\mathcal{N}'}\right)$ is given by

$$R_1\left(\tau_{\pi(i+1)}, \tau_{\pi(i)}|\bar{\gamma}^\star_{\mathcal{N}'}\right) = \int_{\tau_{\pi(i+1)}}^{\infty} \log(1+x)dF^2(x) + F\left(\tau_{\pi(i+1)}\right)\int_{\tau_{\pi(i)}}^{\tau_{\pi(i+1)}} \log(1+x)dF(x)$$
$$+ \log\left(1+\bar{\gamma}^\star_{\mathcal{N}'}\right)F\left(\tau_{\pi(i)}\right)F\left(\tau_{\pi(i+1)}\right).$$

*Proof:* See Appendix D. ■

Finally, we look at the case where $\tau_{\pi(i)} \le \bar{\gamma}^\star_{\mathcal{N}'} \le \tau_{\pi(i+1)}$.

*Lemma 8:* If $\tau_{\pi(i)} \le \bar{\gamma}^\star_{\mathcal{N}'} \le \tau_{\pi(i+1)}$, $R^1\left(\tau_{\pi(i+1)}, \tau_{\pi(i)}|\bar{\gamma}^\star_{\mathcal{N}'}\right)$ is given by

$$R_2\left(\tau_{\pi(i+1)}, \tau_{\pi(i)}|\bar{\gamma}^\star_{\mathcal{N}'}\right) = \int_{\tau_{\pi(i+1)}}^{\infty} \log(1+x)dF^2(x) + F\left(\tau_{\pi(i+1)}\right)\int_{\bar{\gamma}^\star_{\mathcal{N}'}}^{\tau_{\pi(i+1)}} \log(1+x)dF(x)$$
$$+ \log\left(1+\bar{\gamma}^\star_{\mathcal{N}'}\right)F\left(\tau_{\pi(i+1)}\right)F\left(\bar{\gamma}^\star_{\mathcal{N}'}\right).$$

*Proof:* See Appendix D. ■

For the final two cases, we note that $R^1\left(\tau_{\pi(i+1)}, \tau_{\pi(i)}|\bar{\gamma}^\star_{\mathcal{N}'}\right)$ depends both on threshold values $\tau_{\pi(i)}$ and $\tau_{\pi(i+1)}$, and on $\bar{\gamma}^\star_{\mathcal{N}'}$. The results of these three lemmas have been graphically summarized in Fig. 2.

If $\bar{\gamma}^\star_{\mathcal{N}'} = \tau_{\pi(i)}$, $R_1$ and $R_2$ in Lemmas 7 and 8 evaluate to the same expression. Similarly, if $\bar{\gamma}^\star_{\mathcal{N}'} = \tau_{\pi(i+1)}$, $R_0$ and $R_2$ in Lemmas 6 and 8 evaluate to the same expression. This shows that the rate as a function of $\bar{\gamma}^\star_{\mathcal{N}'}$ is continuous at $\tau_{\pi(i)}$ and $\tau_{\pi(i+1)}$.

Given the initial threshold values $\left\{\tau_{\pi(k)}\right\}_{k=1}^{n}$, the first step to discover the Schur-concave structure in the sum-rate function is to analyze the behavior of the function

$$g_T(\epsilon) = R^1\left(\tau_{\pi(i+1)} + \epsilon, \tau_{\pi(i)} - \epsilon | \bar{\gamma}^\star_{\mathcal{N}'}\right)$$

for $\epsilon \in \left[0, \min\left\{\tau_{\pi(i)} - \tau_{\pi(i-1)}, \tau_{\pi(i+2)} - \tau_{\pi(i+1)}\right\}\right]$ by making use of Lemma 4. This is now a scalar problem. At this point, it is more useful to interpret the sum-rate as a function of feedback probabilities since the feedback constraint in (14) is in terms of these probabilities. This interpretation helps us to incorporate the feedback constraints into our optimization problem more easily, as will be shown next.

*2) Rate as a Function of Feedback Probabilities:* There is a one-to-one correspondence between feedback thresholds $\tau_{\pi(i)}$ and feedback probabilities $p_{\pi(i)}$ since $f$ has the support $\mathbb{R}_+$, *i.e.*, $\tau_{\pi(i)} = F^{-1}(1 - p_{\pi(i)})$. Hence, we can represent $R^1\left(\tau_{\pi(i+1)}, \tau_{\pi(i)} | \bar{\gamma}^\star_{\mathcal{N}'}\right)$ as $R^1\left(p_{\pi(i)}, p_{\pi(i+1)} | \bar{\gamma}^\star_{\mathcal{N}'}\right)$ without any ambiguity. With this interpretation, the optimization problem in (14) can be considered as the problem of finding optimum feedback probability vector $\boldsymbol{p}^\star = \left(p_1^\star, \ldots, p_n^\star\right)^\top$ in $[0,1]^n$ subject to the feedback constraint $\sum_{i=1}^n p_i^\star \leq \lambda$. Indeed, it is easy to see that any feedback policy solving (14) must achieve the feedback constraint with equality, *i.e.*, $\sum_{i=1}^n p_i^\star = \lambda$.

Since $F$ is monotone increasing, we have $p_{\pi(1)} \geq p_{\pi(2)} \geq \cdots \geq p_{\pi(i)} \geq p_{\pi(i+1)} \geq \cdots \geq p_{\pi(n)}$. Focusing on $p_{\pi(i)}$ and $p_{\pi(i+1)}$, we have the feedback level $\lambda_i = p_{\pi(i)} + p_{\pi(i+1)}$, and other probabilities give us natural boundaries on $p_{\pi(i)}$ and $p_{\pi(i+1)}$ as such $p_{\pi(i+2)} \leq p_{\pi(i+1)} \leq p_{\pi(i)} \leq p_{\pi(i-1)}$. Without violating these boundaries, we will vary $p_{\pi(i)}$ and $p_{\pi(i+1)}$ by keeping $\lambda_i$ constant.

Similar to the previous part, we start our analysis by focusing on a two-user system. Given a feedback constraint $\lambda > 0$, we can restrict our search for the optimal feedback probability vector to the plane given by $p_{\pi(1)} + p_{\pi(2)} = \lambda$. On this plane, we write the rate function $R^1(\boldsymbol{p})$ as a function of only $p_{\pi(2)}$ without any ambiguity. The communication rate on this plane as a function of $p_{\pi(2)}$ is given below.

*Lemma 9:* The rate on beam $1$ of a two-user system on the plane

$$\mathcal{P} = \left\{\boldsymbol{p} \in [0,1]^2 : p_{\pi(1)} + p_{\pi(2)} = \lambda\right\}$$

as a function of $p_{\pi(2)}$ is equal to

$$R^1\left(p_{\pi(2)}\right) = \int_{F^{-1}(1-p_{\pi(2)})}^{\infty} \log\left(1+x\right) dF^2(x) + \left(1 - p_{\pi(2)}\right) \int_{F^{-1}(1+p_{\pi(2)}-\lambda)}^{F^{-1}(1-p_{\pi(2)})} \log\left(1+x\right) dF(x).$$

for $\max\{0, \lambda - 1\} \leq p_{\pi(2)} \leq \frac{\lambda}{2}$

*Proof:* Follows from a direct substitution of $\tau_{\pi(2)} = F^{-1}\left(1 - p_{\pi(2)}\right)$ in Lemma 5. ■

$F^{-1}$ in the expression above represents the functional inverse of $F$. We give the first derivative of the two-user rate in the following lemma.

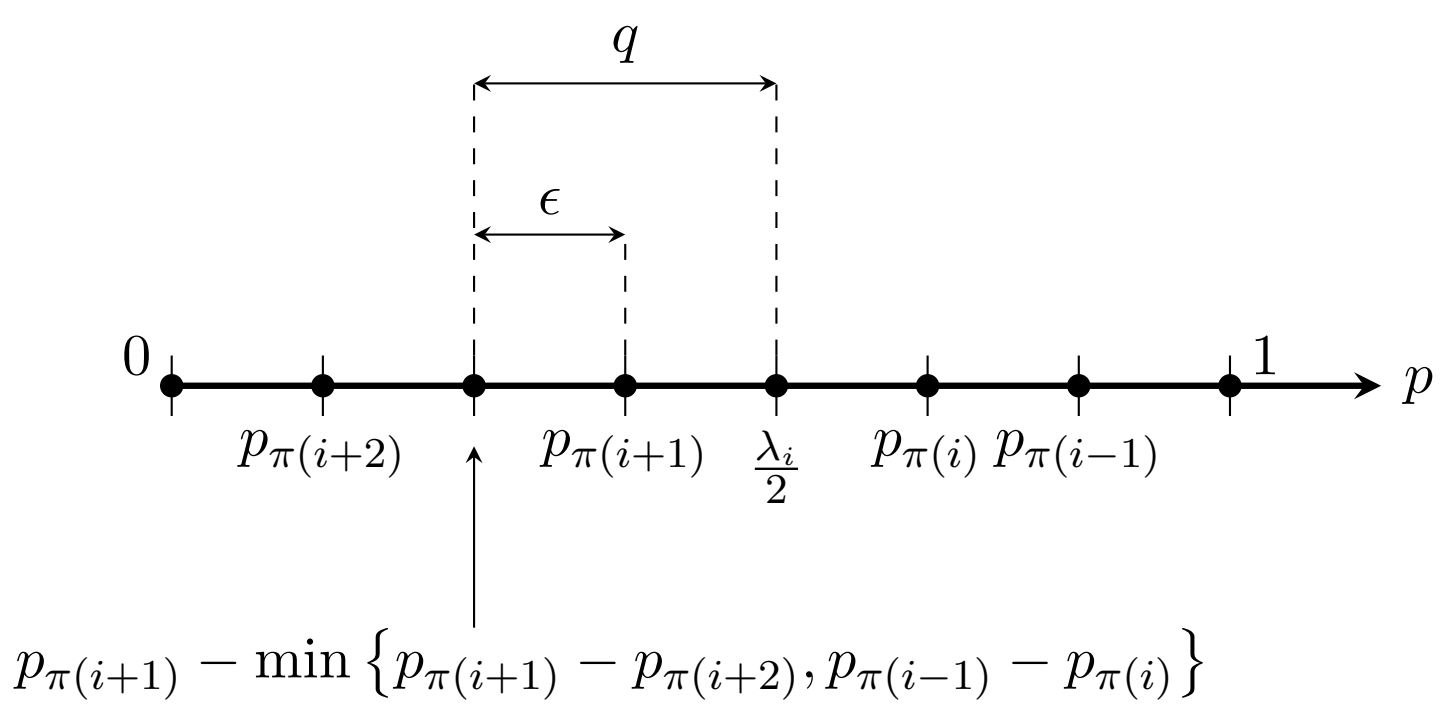

Fig. 3. Ordered feedback probabilities, and the range of $q$ and $\epsilon$.

*Lemma 10:* The first derivative of $R^1\left(p_{\pi(2)}\right)$ on

$$\mathcal{P} = \left\{\boldsymbol{p} \in [0,1]^2 : p_{\pi(1)} + p_{\pi(2)} = \lambda\right\}$$

is equal to

$$\frac{dR^1\left(p_{\pi(2)}\right)}{dp_{\pi(2)}} = \int_{F^{-1}(1+p_{\pi(2)}-\lambda)}^{F^{-1}(1-p_{\pi(2)})} \frac{F(x)}{1+x}dx - \left(\lambda - 2p_{\pi(2)}\right)\log\left(1 + F^{-1}\left(1 + p_{\pi(2)} - \lambda\right)\right). \tag{22}$$

for $\max\{0, \lambda-1\} \leq p_{\pi(2)} \leq \frac{\lambda}{2}$.

*Proof:* Follows directly after differentiating the rate expression in Lemma 9. ■

We note that Lemma 4 implies the necessity of $\frac{dR^1(p_{\pi(2)})}{dp_{\pi(2)}} \geq 0$ for all $p_{\pi(2)} \in \left[\max\{0, \lambda-1\}, \frac{\lambda}{2}\right]$ and $\lambda \in [0,2]$ for the Schur-concavity of the two-user sum-rate. Consider now the $n$-user scenario. Given the initial feedback probabilities $\left\{p_{\pi(k)}\right\}_{k=1}^{n}$, we need to analyze the behavior of the function

$$g_p(\epsilon) = R^1\left(p_{\pi(i)} + \epsilon, p_{\pi(i+1)} - \epsilon | \bar{\gamma}^{\star}_{\mathcal{N}'}\right) \tag{23}$$

for $\epsilon \in \left[0, \min\left\{p_{\pi(i-1)} - p_{\pi(i)}, p_{\pi(i+1)} - p_{\pi(i+2)}\right\}\right]$ to discover Schur-concavity of the rate function by Lemma 4. We have already discussed how we can vary $p_{\pi(2)}$ by keeping $\lambda$ constant for the two-user case. Analysis of the general $n$-user scenario is not fundamentally different from the two-user scenario, and a similar technique used for the analysis of the two-user rate function can still be applied for the general $n$-user case without violating the boundary conditions on feedback probabilities. That is, we introduce an auxiliary variable $q \in \mathcal{P}_{i+1}$, replace $p_{\pi(i+1)} - \epsilon$ with $q$ and $p_{\pi(i)} + \epsilon$ with $\lambda_i - q$, and write $R^1\left(p_{\pi(i)} + \epsilon, p_{\pi(i+1)} - \epsilon | \bar{\gamma}^{\star}_{\mathcal{N}'}\right)$ as a function of $q$, where

$$\mathcal{P}_{i+1} = \left[p_{\pi(i+1)} - \min\left\{p_{\pi(i+1)} - p_{\pi(i+2)}, p_{\pi(i-1)} - p_{\pi(i)}\right\}, \frac{\lambda_i}{2}\right].$$

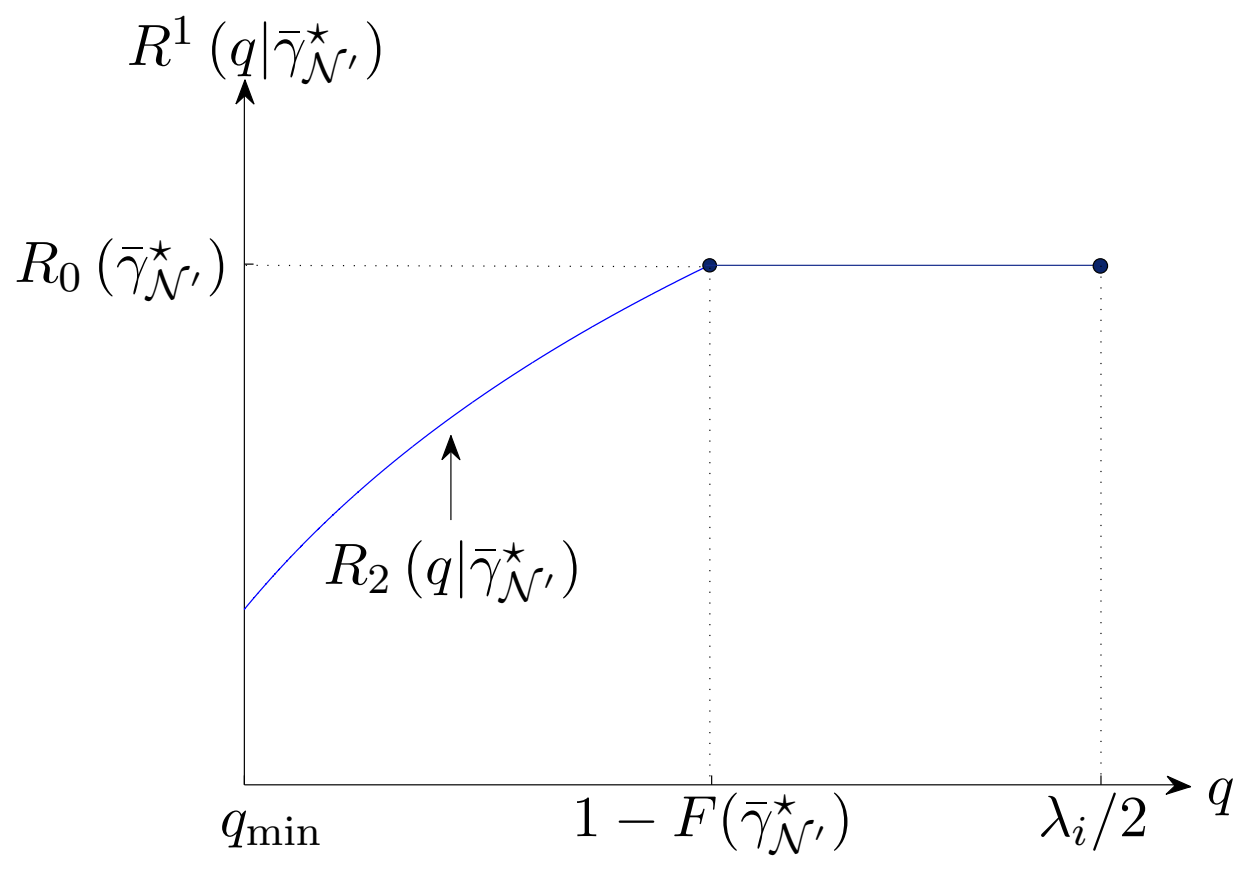

Fig. 4. A pictorial representation for the rate expression in (24) for $q_{\min} \leq 1 - F(\bar{\gamma}^\star_{\mathcal{N}'}) \leq \frac{\lambda_i}{2}$.

Fig. 3 provides a graphical representation for the selection of $q$. By using Lemma 6, 7 and 8, we have

$$R^1(q|\bar{\gamma}^\star_{\mathcal{N}'}) = R_0(\bar{\gamma}^\star_{\mathcal{N}'}) 1_{\left\{q>1-F\left(\bar{\gamma}^\star_{\mathcal{N}'}\right)\right\}} + R_1(q|\bar{\gamma}^\star_{\mathcal{N}'}) 1_{\left\{q>\lambda_i-\left(1-F\left(\bar{\gamma}^\star_{\mathcal{N}'}\right)\right)\right\}} + R_2(q|\bar{\gamma}^\star_{\mathcal{N}'}) 1_{\left\{q\leq 1-F(\bar{\gamma}^\star_{\mathcal{N}'})\ \&\ q\leq\lambda_i-\left(1-F\left(\bar{\gamma}^\star_{\mathcal{N}'}\right)\right)\right\}} \quad (24)$$

for $q \in \mathcal{P}_{i+1}$.

Some insights about (24) are as follows. Let $q_{\min} = p_{\pi(i+1)} - \min\left\{p_{\pi(i+1)} - p_{\pi(i+2)}, p_{\pi(i-1)} - p_{\pi(i)}\right\}$, and assume $1-F(\bar{\gamma}^\star_{\mathcal{N}'}) \leq \frac{\lambda_i}{2}$. If $1-F(\bar{\gamma}^\star_{\mathcal{N}'}) < q_{\min}$, $R^1(q|\bar{\gamma}^\star_{\mathcal{N}'})$ is equal to $R_0(\bar{\gamma}^\star_{\mathcal{N}'})$ for all $q \in \left[q_{\min}, \frac{\lambda_i}{2}\right]$. On the other hand, if $1 - F(\bar{\gamma}^\star_{\mathcal{N}'}) \geq q_{\min}$, $R^1(q|\bar{\gamma}^\star_{\mathcal{N}'})$ first becomes equal to $R_2(q|\bar{\gamma}^\star_{\mathcal{N}'})$ and then equal to $R_0(\bar{\gamma}^\star_{\mathcal{N}'})$ as $q$ changes from $q_{\min}$ to $\frac{\lambda_i}{2}$. This behavior is graphically depicted in Fig. 4.[7] Therefore, the rate in this case can be visualized as a concatenation of two functions with a gluing point at $1 - F(\bar{\gamma}^\star_{\mathcal{N}'})$. Similar explanations can be given for $1 - F(\bar{\gamma}^\star_{\mathcal{N}'}) > \frac{\lambda_i}{2}$.

*3) Schur-concavity of the Sum-rate Function:* Building upon our analysis above, we will obtain sufficient conditions for the Schur-concavity of the sum-rate in this part. We start our analysis by first providing a proof for Theorem 5. We restated Theorem 5 below for the sake of completeness.

*Theorem 5:* The sum-rate $R(\boldsymbol{p})$ is a Schur-concave function if

$$\log(1+\gamma)(\lambda - 2q) + \int_{F^{-1}(1+q-\lambda)}^{F^{-1}(1-q)} \frac{F(x)}{1+x} dx - (\lambda - 2q)\log\left(1 + F^{-1}(1+q-\lambda)\right) \geq 0$$

for all $\gamma \geq 0$, $\lambda \in [0, 2]$ and $\max\{0, \lambda - 1\} \leq q \leq \frac{\lambda}{2}$.

[7]The plot may not be exactly accurate. It is just given to conceptualize the behavior of the rate function.

*Proof:* It is enough to show that $R^1(q|\bar{\gamma}^\star_{\mathcal{N}'})$ is a non-decreasing function of $q \in \mathcal{P}_{i+1}$ for all $i = 1, \ldots, n-1$ and $\bar{\gamma}^\star_{\mathcal{N}'} \geq 0$ based on Lemma 4. To this end, we can write $R_1(q|\bar{\gamma}^\star_{\mathcal{N}'})$ explicitly as

$$R_1(q|\bar{\gamma}^\star_{\mathcal{N}'}) = \int_{F^{-1}(1-q)}^{\infty} \log(1+x)dF^2(x) + (1-q)\int_{F^{-1}(1+q-\lambda_i)}^{F^{-1}(1-q)} \log(1+x)dF(x) + \log(1+\bar{\gamma}^\star_{\mathcal{N}'})(1-q)(1+q-\lambda_i).$$

Using Lemma 10, we get

$$\frac{dR_1(q|\bar{\gamma}^\star_{\mathcal{N}'})}{dq} = \log(1+\bar{\gamma}^\star_{\mathcal{N}'})(\lambda_i - 2q) + \int_{F^{-1}(1+q-\lambda_i)}^{F^{-1}(1-q)} \frac{F(x)}{1+x}dx - (\lambda_i - 2q)\log\left(1+F^{-1}(1+q-\lambda_i)\right). \quad (25)$$

Similarly, we can write $R_2(q|\bar{\gamma}^\star_{\mathcal{N}'})$ explicitly as

$$R_2(q|\bar{\gamma}^\star_{\mathcal{N}'}) = \int_{F^{-1}(1-q)}^{\infty} \log(1+x)dF^2(x) + (1-q)\int_{\bar{\gamma}^\star_{\mathcal{N}'}}^{F^{-1}(1-q)} \log(1+x)dF(x) + \log(1+\bar{\gamma}^\star_{\mathcal{N}'})(1-q)F(\bar{\gamma}^\star_{\mathcal{N}'}).$$

Differentiation and integration-by-parts give us

$$\frac{dR_2(q|\bar{\gamma}^\star_{\mathcal{N}'})}{dq} = \int_{\bar{\gamma}^\star_{\mathcal{N}'}}^{F^{-1}(1-q)} \frac{F(x)}{1+x}dx \geq 0.$$

Thus, $R^1(q|\bar{\gamma}^\star_{\mathcal{N}'})$ is a non-decreasing function of $q \in \mathcal{P}_{i+1}$ for all $i = 1, \ldots, n-1$ and $\bar{\gamma}^\star_{\mathcal{N}'} \geq 0$ if (16) is correct. ■

Second, we provide a proof for Theorem 6 based on Theorem 5. The new sufficient condition for the Schur-concavity of the sum-rate function is obtained by means of a second order analysis. Although complex looking, it turns out to be much easier to deal with analytically as illustrated for Rayleigh fading channels in the next section. Again, we restate Theorem 6 below for the sake of completeness.

*Theorem 6:* The sum-rate $R(\boldsymbol{p})$ is Schur-concave if $f$ is bounded at zero, and has the derivative $f'$ satisfying $f'\left(F^{-1}(x)\right) \leq -\frac{f(F^{-1}(x))}{1+F^{-1}(x)}$ for all $x \in [0,1]$.

*Proof:* Let $U(q,\lambda) = \int_{F^{-1}(1+q-\lambda)}^{F^{-1}(1-q)} \frac{F(x)}{1+x}dx - (\lambda - 2q)\log\left(1+F^{-1}(1+q-\lambda)\right)$. Then, it is enough to show that $U(q,\lambda) \geq 0$ for all $\lambda \in [0,2]$ and $\max\{0, \lambda-1\} \leq q \leq \frac{\lambda}{2}$ by Theorem 5. To this end, it is enough to show $\frac{\partial U(q,\lambda)}{\partial q} \leq 0$ for all $\lambda \in [0,2]$ and $\max\{0,\lambda-1\} \leq q \leq \frac{\lambda}{2}$ since $U\left(\frac{\lambda}{2}, \lambda\right) = 0$.

The following lemma simplifies the proof considerably.

*Lemma 11:* Let $G(x) = \log\left(1+F^{-1}(x)\right)\left(1+F^{-1}(x)\right) f\left(F^{-1}(x)\right) - x$ for $x \in [0,1]$. If $f$ is bounded at zero and $f'$ satisfies $f'\left(F^{-1}(x)\right) \leq -\frac{f(F^{-1}(x))}{1+F^{-1}(x)}$ for all $x \in [0,1]$, then $G \leq 0$ on $[0,1]$.

*Proof:* By taking the first derivative of $G(x)$ with respect to $x$,

$$\begin{aligned}\frac{dG(x)}{dx} &= \log\left(1+F^{-1}(x)\right)\left(1+F^{-1}(x)\right)\frac{f'\left(F^{-1}(x)\right)}{f\left(F^{-1}(x)\right)}+\log\left(1+F^{-1}(x)\right)\\ &= \log\left(1+F^{-1}(x)\right)\left[1+\frac{\left(1+F^{-1}(x)\right)f'\left(F^{-1}(x)\right)}{f\left(F^{-1}(x)\right)}\right]<0.\end{aligned}$$

Hence, $G(x)$ is strictly decreasing for $x>0$, and achieves its maximum at $x=0$. We have $\lim_{x\to 0} G(x)=0$ since $f(x)$ is bounded at $0$, which completes the proof. ■

Now, consider the partial derivative of $U(q,\lambda)$ with respect to $q$, which is equal to

$$\begin{aligned}\frac{\partial U(q,\lambda)}{\partial q} &= \frac{1-q}{1+F^{-1}(1-q)}.\frac{-1}{f\left(F^{-1}(1-q)\right)}-\frac{1+q-\lambda}{1+F^{-1}(1+q-\lambda)}.\frac{1}{f\left(F^{-1}(1+q-\lambda)\right)}\\ &- \frac{\lambda-2q}{1+F^{-1}(1+q-\lambda)}.\frac{1}{f\left(F^{-1}(1+q-\lambda)\right)}+2\log\left(1+F^{-1}(1+q-\lambda)\right).\end{aligned}$$

Taking the common denominators gives us

$$\frac{\partial U(q,\lambda)}{\partial q}=K_1(q)\,g_1(q,\lambda)+K_2(q,\lambda)\,g_2(q,\lambda),$$

where

$$\begin{aligned}g_1(q,\lambda) &= \log\left(1+F^{-1}(1+q-\lambda)\right)\left(1+F^{-1}(1-q)\right)f\left(F^{-1}(1-q)\right)-(1-q),\\ g_2(q,\lambda) &= \log\left(1+F^{-1}(1+q-\lambda)\right)\left(1+F^{-1}(1+q-\lambda)\right)f\left(F^{-1}(1+q-\lambda)\right)-(1-q),\\ K_1(q) &= \frac{1}{\left(1+F^{-1}(1-q)\right)f\left(F^{-1}(1-q)\right)},\ \text{and}\\ K_2(q,\lambda) &= \frac{1}{\left(1+F^{-1}(1+q-\lambda)\right)f\left(F^{-1}(1+q-\lambda)\right)}.\end{aligned}$$

Note that $K_1$ and $K_2$ are always positive. Thus, it is enough to show that $g_1$ and $g_2$ are non-positive on $\left[\max\{0,\lambda-1\},\frac{\lambda}{2}\right]$ for any fixed $\lambda\in[0,2]$. To this end, $g_1$ and $g_2$ on $\left[\max\{0,\lambda-1\},\frac{\lambda}{2}\right]$ can be upper bounded as

$$g_1(q,\lambda)\le g_1^u(q) = \log\left(1+F^{-1}(1-q)\right)\left(1+F^{-1}(1-q)\right)f\left(F^{-1}(1-q)\right)-(1-q)$$

and

$$\begin{aligned}g_2(q,\lambda) &\le g_2^u(q,\lambda)\\ &= \log\left(1+F^{-1}(1+q-\lambda)\right)\left(1+F^{-1}(1+q-\lambda)\right)f\left(F^{-1}(1+q-\lambda)\right)-(1+q-\lambda).\end{aligned}$$

Now, using Lemma 11, we can show that both $g_1^u$ and $g_2^u$ are non-positive functions on $\left[\max(0,\lambda-1),\frac{\lambda}{2}\right]$. This means $\frac{\partial U(q,\lambda)}{\partial q}\le 0$, which implies $U(q,\lambda)\ge 0$ for all $\lambda\in[0,2]$ and $\max\{0,\lambda-1\}\le q\le\frac{\lambda}{2}$. ■

## VI. Applications and Discussion

In this section, we will apply our results derived in Sections IV and V to well known fading channel models. We will also discuss intuition behind the resulting performance figures. We start our discussion with Rayleigh fading channels, which is one of the most commonly used channel models in the literature, *e.g.,* see [32]–[34], and closely approximates measured data rates in densely populated urban areas [35].

### A. Rayleigh Fading Channels

Consider the Rayleigh fading channel model in which $h_{k,i}, k = 1, \ldots, N_t$ and $i = 1, \ldots, n$, are assumed to be i.i.d. with the common distribution $\mathcal{CN}(0,1)$, where $\mathcal{CN}\left(\mu, \sigma^2\right)$ represents the *circularly-symmetric complex Gaussian* distribution with mean $\mu$ and variance $\sigma^2$. Recall that the background noise is the unit power (complex) Gaussian noise, and therefore $\rho$ is interpreted as the average $\mathrm{SNR}$ below.

For this channel model, the $\mathrm{SINR}$ distribution function $F$ and the associated probability density function $f$ can be given as

$$F(x) = 1 - \frac{e^{-\frac{x}{\rho}}}{(x+1)^{M-1}} \tag{26}$$

and

$$f(x) = \frac{e^{-\frac{x}{\rho}}}{(x+1)^{M}} \left[\frac{1}{\rho}(x+1) + M - 1\right], \tag{27}$$

respectively [13]. An important quantity of interest to apply our results in Theorems 5 and 6 is the functional inverse, $F^{-1}$, of $F$. The next lemma provides an analytical expression for $F^{-1}$ for Rayleigh fading channels.

*Lemma 12:* $F^{-1}$ is equal to

$$F^{-1}(x) = \begin{cases} -1 + (M-1)\rho W\left(\frac{\exp\left(\frac{1}{(M-1)\rho}\right)}{(M-1)\rho}(1-x)^{\frac{1}{1-M}}\right) & \text{if } M \geq 2 \\ -\rho \log(1-x) & \text{if } M = 1 \end{cases},$$

where $x \in [0,1]$ and $W$ is the Lambert W function given by the defining equation $W(x)\exp(W(x)) = x$ for $x \geq -\frac{1}{e}$.

*Proof:* See Appendix E. ∎

To motivate the discussion below, we start by providing two simple numerical examples; first of which illustrates a network configuration in which homogenous threshold feedback policies are optimal, whereas the second example provides another network configuration in which homogenous threshold feedback policies are strictly suboptimal.

Consider two MUs located in a Rayleigh fading environment, *i.e.*, all channel (amplitude) gains are random with distribution $\mathcal{CN}(0,1)$. $M$ and $\lambda$ are chosen to be $M=1$ and $\lambda=0.5$ in both examples below. We set $\rho$ to $0$ [dB] in the first example, while it is set to $10$ [dB] in the second one. Since all MUs are identical with identical fading characteristics in this set-up, it is intuitively expected that a homogenous threshold feedback policy must be optimal, and solve the rate maximization problem in (5) under both network configurations.

This is indeed correct for the first network configuration as shown in Fig 5(a). The sum-rate is clearly maximized at $\boldsymbol{p}=(0.25,0.25)^\top$, and therefore the homogenous threshold feedback policy with thresholds set as $\boldsymbol{\tau}_{\rm homo}=(\log(4),\log(4))^\top$ solves (5). However, this intuition does not always work as illustrated by the second example. In this case, the homogenous threshold feedback policy equalizing the feedback probabilities of MUs becomes strictly suboptimal, *i.e.*, see Fig. 5(b). This shows that $R(\boldsymbol{p})$ is not a Schur-concave function of feedback probabilities for these selections of model parameters, and hence, it is not necessarily maximized at $\boldsymbol{p}=(0.25,0.25)^\top$. We note that the selection of parameters in both examples is just for elucidatory purposes, and the same arguments continue to hold for other values of $\lambda$.

This discussion motivates the following question: When are homogenous threshold feedback policies optimal for Rayleigh fading channels? The answer is supplied by the following two theorems.

*Theorem 7:* For Rayleigh fading environments with $M=1$ and $\rho\leq 1$, $R(\boldsymbol{p})$ is a Schur-concave function of feedback probabilities, and therefore the homogenous threshold feedback policy satisfying feedback constraints with equality solves (5) when $M=1$ and $\rho\leq 1$.

*Theorem 8:* For Rayleigh fading environments with $M>1$, $R(\boldsymbol{p})$ is a Schur-concave function of feedback probabilities, and therefore the homogenous threshold feedback policy satisfying feedback constraints with equality solves (5) when $M>1$.

*Proof:* See Appendix F. ■

Since the proofs are similar, and are based on the sufficient condition established in Theorem 6, we skip the proof of Theorem 7 to avoid repetitions. Theorem 7 shows that it is enough to have $\rho$ smaller than or equal to 1 to ensure the optimality of homogenous threshold feedback policies for Rayleigh fading environments when only a single beam is used for the downlink communication. Since $F$ in (26) does not depend on $N_t$, the same result continues to hold for $N_t>1$ as long as multiple transmit antennas are used to form a single beam as in [5].

On the other hand, Theorem 8 provides an extension of Theorem 7 to multiple beams. Theorem 8 is promising for multiuser MIMO downlink communication in a Rayleigh fading environment because

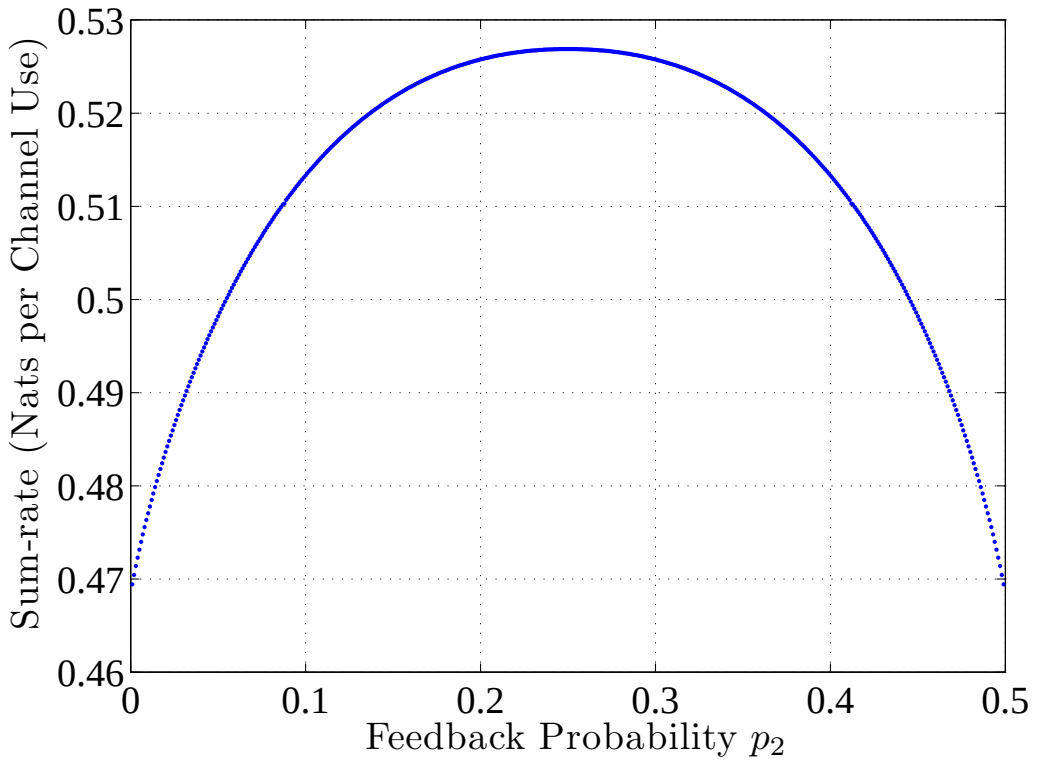

(a) Behavior of the sum-rate as a function of the feedback probability $p_2$ of the second MU for the first example. ($M = 1, \lambda = 0.5$ and $\rho = 0$ [dB])

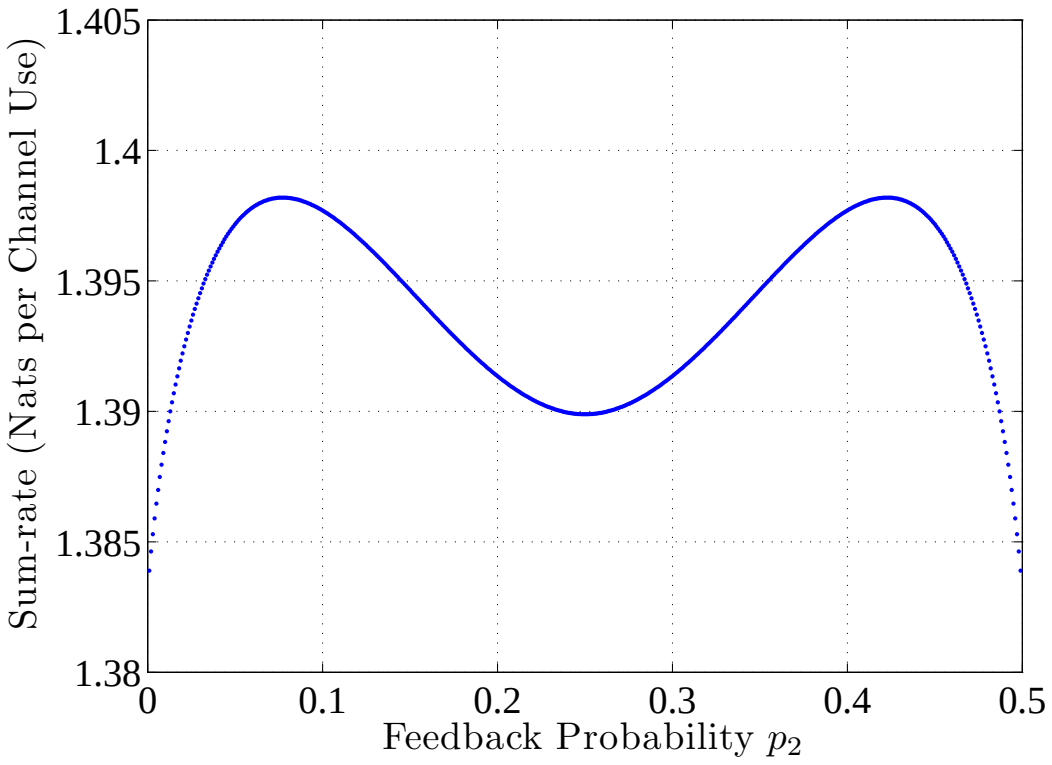

(b) Behavior of the sum-rate as a function of the feedback probability $p_2$ of the second MU for the second example. ($M = 1, \lambda = 0.5$ and $\rho = 10$ [dB]) it shows that homogenous threshold feedback policies are always optimal if multiple beams are used to communicate with multiple MUs simultaneously. Although the optimality of homogenous threshold feedback policies strongly depends on the properties of the underlying fading process modulating received signal strengths and the background noise level present in the system for the single beam case, this is not true anymore for multiple beams. More intuition is provided on this point later.

Fig. 5. Numerical examples illustrating the optimality and sub-optimality of homogenous threshold policies for different network configurations.

From a theoretical viewpoint, it is surprising to see that a property holding in the setting of a more complicated and general MIMO system model does not always hold for single-input systems. From a practical viewpoint, MIMO technology is becoming an integral part and a key feature of the next generation wireless communication systems. Thus, these results provide analytically justified design guidelines to maximize data rates subject to feedback constraints in densely populated urban areas with 4G communication systems.

In the second example above, the rate loss due to use of the homogenous feedback policy seems to be very minor around $0.01$ [nats per channel use], and therefore it can be thought to be negligible for all practical purposes. This motivates us to examine the rate difference between homogenous and optimal threshold feedback policies for a broad spectrum of the $\mathsf{SNR}$ parameter to verify or falsify the validity of this conception. To this end, we investigate the optimality gap arising from the use of homogenous threshold feedback policies as opposed to choosing thresholds optimally to maximize the sum-rate in

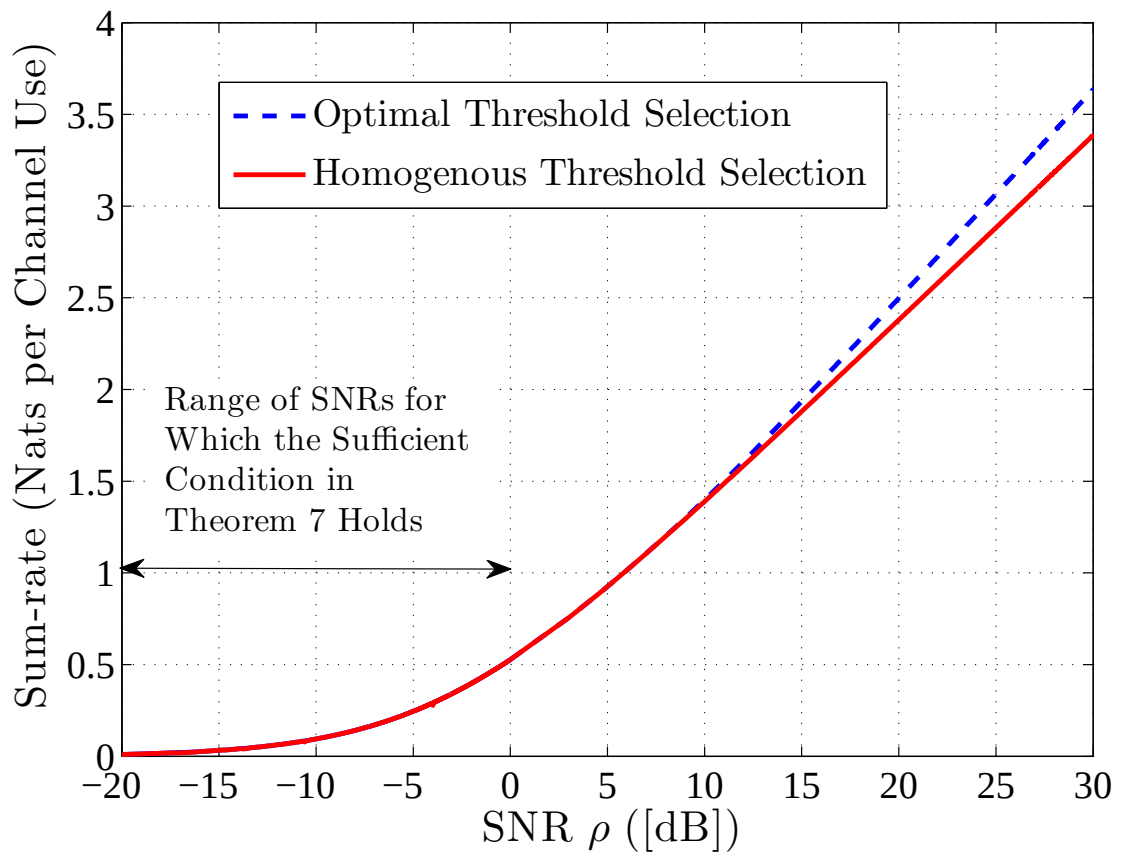

Fig. 6. The optimality gap arising from the use of homogenous threshold feedback policies for different values of $\rho$ when $M = 1$.

Rayleigh fading environments in Fig. 6. We set $n$ to 2, $\lambda$ to 0.5 and $M$ to 1 in this numerical example. Note that homogenous threshold feedback policies are always optimal when $M > 1$. Hence, there is no optimality gap to investigate in this case. For other values of $\lambda$ and $n$, qualitatively similar observations continue to hold. Since we find optimal threshold levels through an exhaustive search, setting $n$ to 2 limits our search space.

For small values of $\rho$ up to 0 [dB], the homogenous threshold feedback policy with threshold levels set as $\boldsymbol{\tau}_{\text{homo}} = \left(\rho \log\left(4\right), \rho \log\left(4\right)\right)^\top$ is optimum as predicted by Theorem 7. It continues to be optimum for a little while up to around 5.7 [dB] SNR values, and after which it becomes strictly suboptimal to use the homogenous threshold feedback policy in terms of the achieved downlink sum-rate. Furthermore, as channel conditions become better, *i.e.*, large values of $\rho$, the optimality gap becomes larger. Practically, this observation indicates that the use of homogenous threshold feedback policies may lead to excessive rate loss in the high SNR regime for single beam systems when compared to the rate achieved by the optimal feedback policy.

Another important issue to investigate is the amount of feedback reduction that can be achieved by setting thresholds optimally. In Fig. 7, we plot the ratio $\frac{R(\boldsymbol{\tau}^\star)}{R(\mathbf{0})}$ between the rates achieved with and without thresholding as a function of $\lambda$ for different numbers of MUs. In this figure, we set $M$ to 1, and $\rho$ to 1. Again, similar observations continue to hold for other values of $M$ and $\rho$. Since $\rho = 1$, the homogenous threshold feedback policy with thresholds set as $\boldsymbol{\tau}^\star = \left(\rho \log\left(\frac{n}{\lambda}\right), \ldots, \rho \log\left(\frac{n}{\lambda}\right)\right)$ is optimum, *i.e.*, see Lemma 12 and Theorem 7. After inspecting the figure, we see that there is almost no rate loss if the average number of MUs feeding back per beam is around five. We call this critical feedback level $\lambda_{\mathrm{c}}$, which is an important design parameter to be inputed to the higher MAC layer. It is interesting to see that the same design parameter applies to all $\frac{R(\boldsymbol{\tau}^\star)}{R(\mathbf{0})}$ curves that shift to the right only slightly and converge pointwise to a limiting curve as the number of MUs in the system increases.

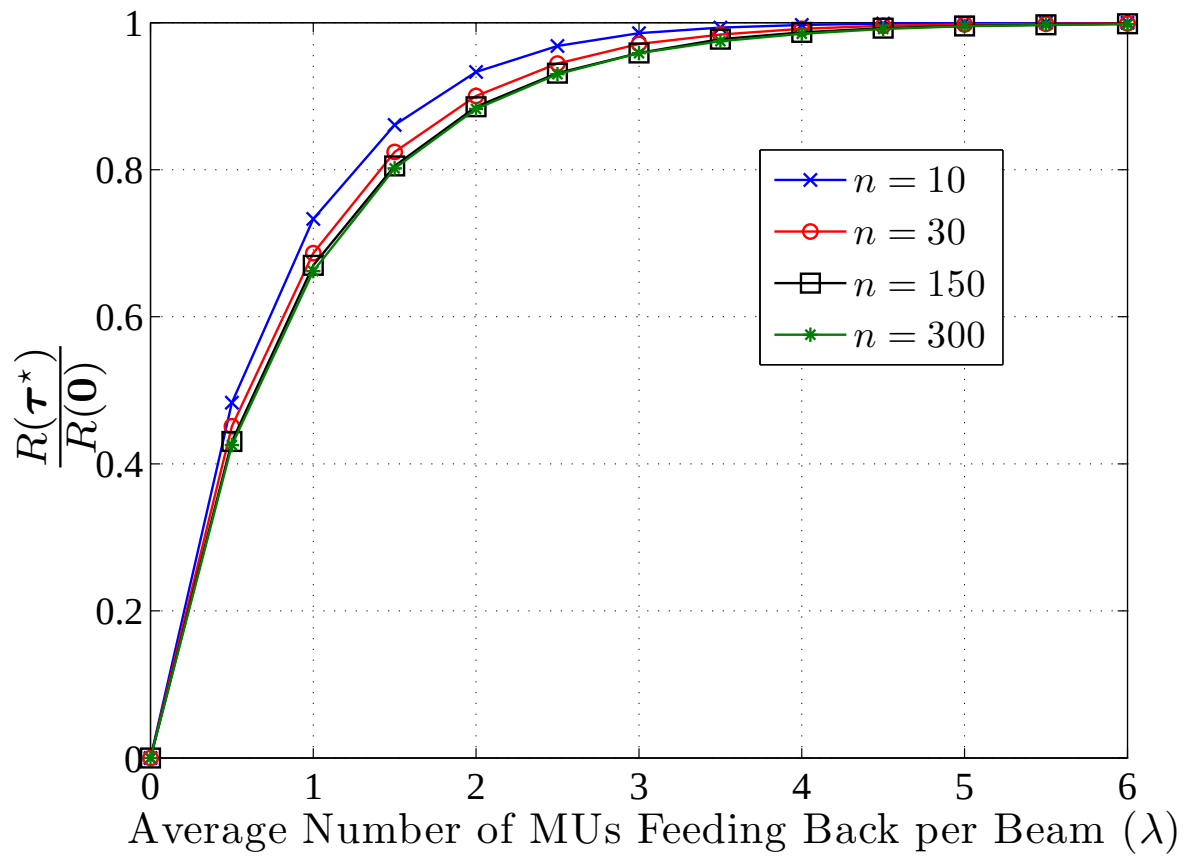

Fig. 7. The ratio between the rates achieved with and without thresholding as a function of the average number of users feeding back per beam for different numbers of MUs.

The reason behind this phenomenon can be explained as follows. No feedback outage event occurs and beams are always assigned to the best MUs at each fading state when thresholds are set to zero. On the other hand, the feedback outage event probability is strictly positive when thresholds are optimally set to meet the feedback constraint $\lambda$. However, the tails of the distribution of the random number of MUs requesting each beam decays to zero exponentially fast, and therefore we are *almost* always guaranteed to have at least one MU demanding each beam whenever $\lambda$ is above the critical feedback level $\lambda_{\mathrm{c}}$. As a result, the feedback outage event probability becomes negligible, and the beams are still assigned to the best MUs with very high probability whenever $\lambda \geq \lambda_{\mathrm{c}}$. Moreover, the distribution of the random number of MUs feeding back converges to a limiting distribution linearly with the total number of MUs in the system, which results in the observed pointwise convergence behavior in Fig. 7. Further details about the limiting $\frac{R(\boldsymbol{\tau}^\star)}{R(\mathbf{0})}$ curve (as $n \to \infty$) can be found in [18], where its exact characterization was obtained and interpreted as the *feedback-capacity tradeoff curve*.

Two possible interpretations about $\lambda_c$ are as follows. Since the BS communicates only with the best MU on each beam, an ideal feedback policy in terms of the optimal usage of uplink communication resources is the one that only allows the best MU to feed back at each channel fading state. However, such a policy requires centralized operation, or coordination among MUs. Thus, when compared with the ideal feedback policy, $\lambda_c$ can be interpreted as the price that we have to pay to achieve *almost* the same performance with the ideal feedback policy due to decentralized operation. Secondly, when compared with the all-feedback policy, it represents the amount of feedback reduction that can be achieved without any noticeable performance degradation. For example, as opposed to allowing all MUs to feed back, we can reduce the total feedback load $30$ times and $60$ times by setting thresholds optimally when $n = 150$ and $n = 300$, respectively, without any evident performance loss.

### *B. Rician and Nakagami Fading Channels*

In this part, we extend our analysis above to other channel models by briefly studying optimality and sub-optimality regions for homogenous threshold feedback policies for Nakagami and Rician fading channel models. We set $M$ to $1$ for simplicity. Otherwise, calculations for the $M > 1$ case easily gets very complicated for these channel models, which hinders the intuitive understanding of the results below. In particular, the derivation of the $\mathrm{SINR}$ distribution in the general case becomes very complex.

We start our discussion with Nakagami fading channels. In this case, $h_{k,i}, k = 1, \ldots, N_t$ and $i = 1, \ldots, n$, are i.i.d. with the common distribution $\mathrm{Nakagami}\,(\mu, \omega)$, where $\mu$ and $\omega$ are shape and spread parameters, respectively. Hence, channel power gains are Gamma distributed with distribution $\mathrm{Gamma}\left(\mu, \frac{\omega}{\mu}\right)$, where $\mu$ and $\frac{\omega}{\mu}$ are shape and scale parameters of the associated Gamma distribution, respectively. We first note that $\omega$ is equal to the average channel power gain, and therefore it is set to 1 to be consistent with the Rayleigh fading channel model above. Secondly, if $X$ is a random variable with distribution $\mathrm{Gamma}\left(\mu, \frac{1}{\mu}\right)$, then $aX$ is distributed according to $\mathrm{Gamma}\left(\mu, \frac{a}{\mu}\right)$, where $a$ is a positive real number. Therefore, under this channel model, the $\mathrm{SINR}$[8] distribution is equal to $\mathrm{Gamma}\left(\mu, \frac{\rho}{\mu}\right)$.

In Fig. 8, we illustrate the regions on which homogenous threshold feedback policies are optimal and suboptimal for the Nakagami fading channel model. We set $n$ to $2$ and $\lambda$ to $0.5$ in this figure. The same observations continue hold for other parameter selections. The blue region is computed numerically by using the sufficient condition for the Schur-concavity of the sum-rate in Theorem 5, whereas the red region is obtained by evaluating the sufficient condition in Theorem 6 numerically. As mentioned earlier, the sufficient condition in Theorem 5 is stronger than the one in Theorem 6, which is why the red region is contained within the blue region in Fig. 8. Note that the Nakagami fading model reduces to the Rayleigh fading model, and the red region only covers $\mathrm{SNR}$ values less than one when $\mu = 1$, which is in accordance with our discussion and Theorem 7 above. Surprisingly, our numerical investigation shows that homogenous threshold feedback policies are suboptimal outside the blue region in Fig. 8. Therefore, we conjecture that the condition provided in Theorem 5 is also necessary for the optimality of homogenous threshold feedback policies.

[8] No inter-beam interference exists in the $M = 1$ case. Hence, the random $\mathrm{SINR}$ is the same quantity with the random $\mathrm{SNR}$. We continue to use the term $\mathrm{SINR}$ for this case to avoid any confusion with the average $\mathrm{SNR}$ $\rho$.

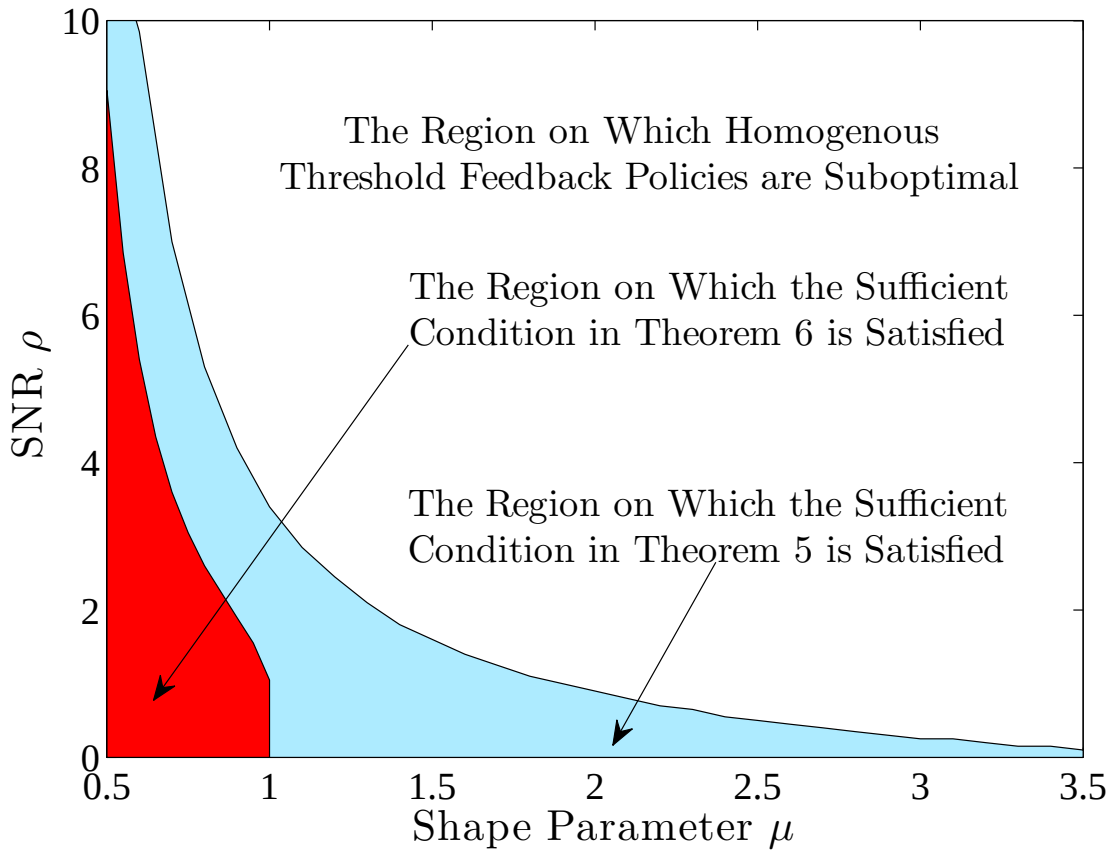

Fig. 8. The regions on which homogenous threshold feedback policies are optimal and suboptimal for the Nakagami fading channel model.

Secondly, we consider the Rician fading channel model in which the channel amplitude gains are Rician distributed with distribution $\text{Rician}\,(K, P)$, where $P$ is the total power gain and $K$ (*a.k.a.*, $K$ factor) is the ratio between the power in the direct path and the power in the scattered paths. We set $P$ to $1$ to be consistent with the Rayleigh and Nakagami fading channel models studied above. If $X$ is a random variable with distribution $\text{Rician}\,(K, P)$, then $\left(\frac{X}{\sigma}\right)^2$ has a non-central Chi-square distribution with two degrees of freedom, and the non-centrality parameter is given by $2K$ if the scaling coefficient $\sigma$ is chosen to be $\sigma = \sqrt{\frac{P}{2(1+K)}}$. We obtain the $\text{SINR}$ distribution by scaling this non-central Chi-square distribution with $\rho\sigma^2$.

Fig. 9 illustrates the regions on which homogenous threshold feedback policies are optimal and suboptimal for the Rician fading channel model. We set $n$ to $2$ and $\lambda$ to $0.5$ in this figure. Since the similar explanations above continue to hold for the Rician fading channel model as well, we do not repeat them here again.

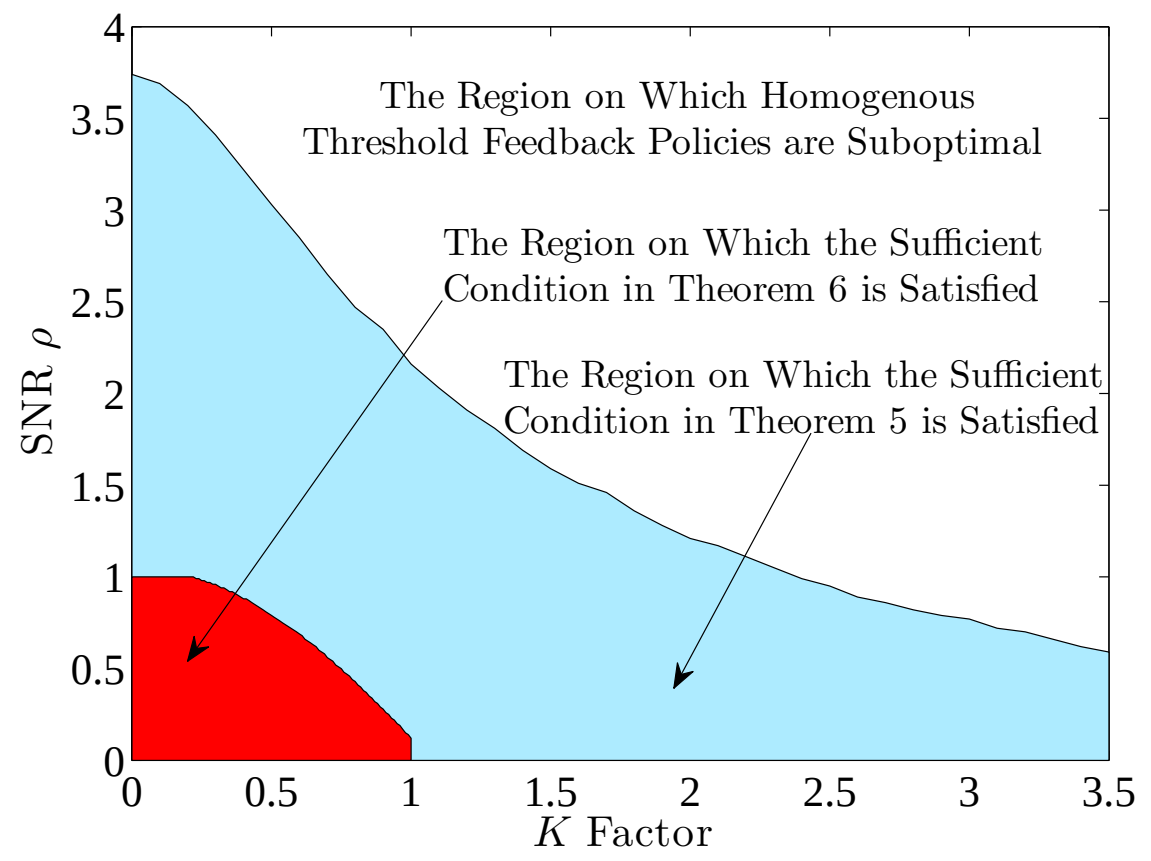

Fig. 9. The regions on which homogenous threshold feedback policies are optimal and suboptimal for the Rician fading channel model.

### C. Why Does Sub-optimality Arise?

In this part, we provide an intuitive explanation for why homogenous threshold feedback policies sometimes become suboptimal to use even when MUs experience statistically the same channel conditions. Our discussion will focus on the single beam case first.

Let $\beta$ be the feedback outage event probability, $R\left(\boldsymbol{\tau}_{\text{homo}}\right)$ be the sum-rate achieved by the homogenous threshold feedback policy satisfying feedback constraints with equality, and $R\left(\boldsymbol{\tau}^{\star}\right)$ be the sum-rate achieved by setting thresholds optimally. For simplicity, we let $n = 2$, but similar explanations continue to hold for any $n$. The sum-rate in this case can be written as

$$R\left(\tau_1, \tau_2\right) = \left(1 - \beta\right) \mathsf{E}\left[\log\left(1 + \max_{i=1,2} \gamma_i \mathbb{1}_{\{\gamma_i \geq \tau_i\}}\right) \middle| \text{ No Outage } \right].$$

Two key underlying factors affect this rate expression. The first one is the *power gain* that can be achieved by means of multiuser diversity. This is represented by the maximization operation inside the logarithm function above. The more MUs feed back, the more likely the output of this maximization operation to be higher. Indeed, the exact asymptotic statistics of the resulting power gain (under various channel models) can be obtained by resorting to an order statistics analysis [36]. The second factor is the *degrees-of-freedom gain* represented by the $1-\beta$ term. The smaller the feedback outage event probability, the higher the degrees-of-freedom gain that we achieve. The choice of thresholds affects both gains, and the interplay between them determines how we set thresholds to maximize the downlink sum-rate.

In Fig. 10, we focus on the Rayleigh fading channel model to provide further details about the interplay between power and degrees-of-freedom gains. In this figure, we set $\lambda$ to $0.5$, and plot the optimal feedback probability $p_2^\star$ of the second MU as a function of $\rho$. In the low $\mathrm{SNR}$ regime, $p_2^\star$ is equal to $0.25$, which implies the optimality of the homogenous threshold feedback policy equalizing the feedback probabilities of both MUs. However, as $\rho$ increases, we start to prefer one MU over the other one to maximize the sum-rate. In this case, for example, we prefer the first MU over the second one by decreasing the feedback probability of the second MU to zero, and increasing the feedback probability of the first MU to $0.5$ in the high $\mathrm{SNR}$ regime.

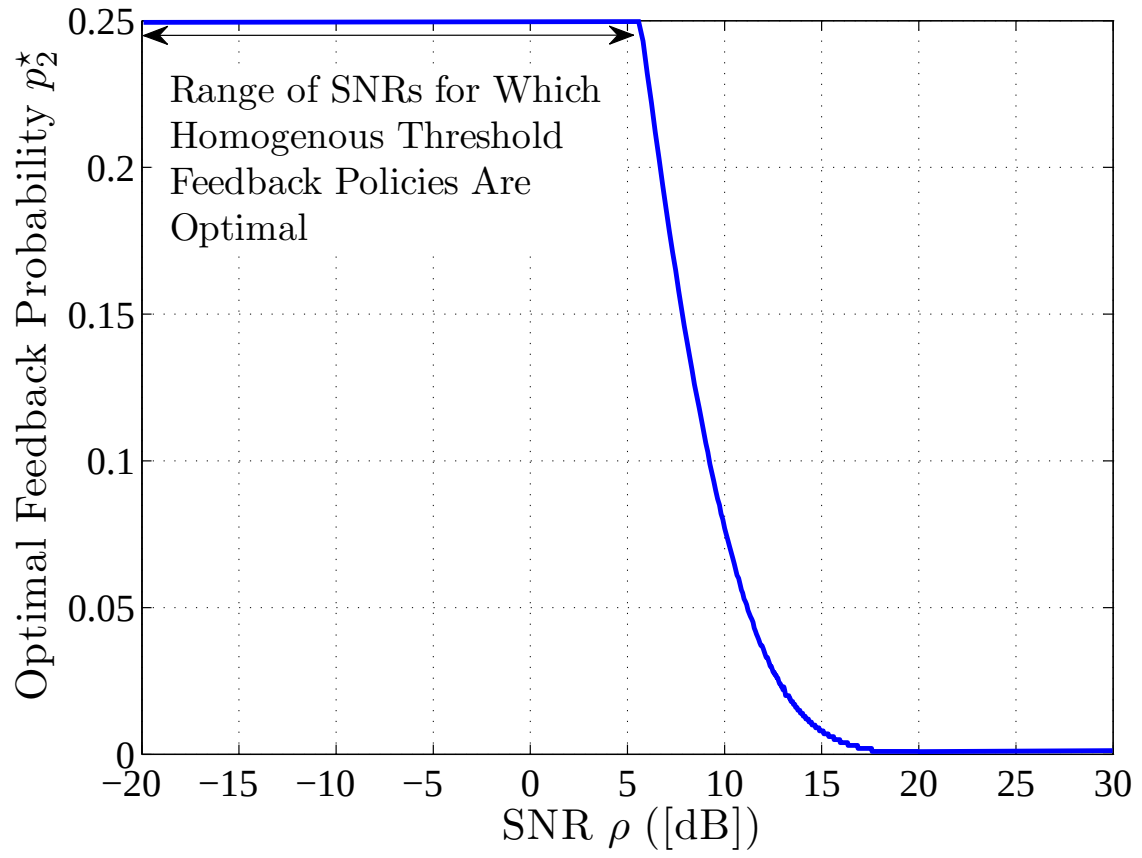

Fig. 10. Optimal feedback probability $p_2^\star$ of the second MU as a function of $\rho$. ($\lambda = 0.5$)

The main reason behind this behavior is as follows. When the $\mathrm{SNR}$ is low, the sum-rate increases *almost* linearly with the power gain. As a result, we tend to choose thresholds equally to maximize the power gain, and thereby to maximize the sum-rate, in the low $\mathrm{SNR}$ regime although such a threshold assignment reduces the degrees-of-freedom gain. In the high $\mathrm{SNR}$ regime, on the other hand, the power gain can only provide a logarithmic increase in the sum-rate, *i.e.*, the law of diminishing returns. Hence, the power gain earned by setting thresholds equally becomes negligible when compared to the loss in the degrees-of-freedom gain, and we tend to choose thresholds heterogeneously to maximize the degrees-of-freedom gain, and thereby to maximize the sum-rate, in the high $\mathrm{SNR}$ regime. A similar behavior continues to hold for other channel models, which is what we investigate next.

In Figs. 11(a) and 11(b), we plot the ratio $\frac{R(\boldsymbol{\tau}_{\mathrm{homo}})}{R(\boldsymbol{\tau}^\star)}$ as a function of $\rho$ and $K$, respectively, for the Rician fading channel model. We set $\lambda$ to $1$ in both figures. The $\mathrm{SNR}$ has the same effect on how we set thresholds optimally in the Rician case as well. When small, we prefer the power gain over the degrees-of-freedom gain, and set thresholds equally to maximize the sum-rate, which is why $\frac{R(\boldsymbol{\tau}_{\mathrm{homo}})}{R(\boldsymbol{\tau}^\star)}$ ratio is around one for small values of $\rho$, and for $K = 0$ and 2. When high, we prefer the degrees-of-freedom gain over the power gain, and set thresholds unequally to maximize the sum-rate, which is why $\frac{R(\boldsymbol{\tau}_{\text{homo}})}{R(\boldsymbol{\tau}^{\star})}$ ratio converges to $0.75$ for high values of $\rho$.

The exact behavior of $\frac{R(\boldsymbol{\tau}_{\text{homo}})}{R(\boldsymbol{\tau}^{\star})}$ strongly depends on $K$, too. Roughly speaking, $K$ determines the dynamic range of the $\mathrm{SINR}$ distribution, and the power gain due to multiuser diversity becomes more prominent when the dynamic range of the distribution is large [4]. However, as $K$ increases, the power in the direct path increases, which, in turn, nullifies the scattering effects and reduces the dynamic range of the $\mathrm{SINR}$ distribution, *e.g.,* see Fig. 12 for an illustration. Therefore, regardless of how small the $\mathrm{SNR}$ is, it may still become suboptimal to use homogenous threshold feedback policies when $K$ is large, as illustrated by the curves corresponding to $K = 10$ and $50$ in Fig. 11(a). Furthermore, as $K$ increases, the channel becomes more deterministic, and we experience almost no power gain due to multiuser diversity in the limit. As a result, $\frac{R(\boldsymbol{\tau}_{\text{homo}})}{R(\boldsymbol{\tau}^{\star})}$ still converges to $0.75$ as $K$ grows large, which is illustrated by Fig. 11(b).

Finally, we note that the limiting value of $\frac{R(\boldsymbol{\tau}_{\text{homo}})}{R(\boldsymbol{\tau}^{\star})}$ (in the high $\mathrm{SNR}$, or high $K$ regime) depends on the feedback constraint $\lambda$. If $\lambda \leq 1$, the optimum feedback probability selection converges to $p_1 = \lambda$ and $p_2 = 0$ (or, vice versa) when $\rho$ or $K$ grows large. Hence, $\frac{R(\boldsymbol{\tau}_{\text{homo}})}{R(\boldsymbol{\tau}^{\star})}$ converges to $1 - \frac{\lambda}{4}$ for $\lambda \leq 1$, which is inline with the $0.75$ limit to which the curves in both Figs. 11(a) and 11(b) converge. If $\lambda > 1$, the optimum feedback probability selection converges to $p_1 = 1$ and $p_2 = \lambda - 1$ (or, vice versa) when $\rho$ or $K$ grows large. Hence, $\frac{R(\boldsymbol{\tau}_{\text{homo}})}{R(\boldsymbol{\tau}^{\star})}$ converges to $\lambda - \frac{\lambda^2}{4}$ for $\lambda > 1$. Let $C^{\star}(\lambda)$, $0 \leq \lambda \leq 2$, be the limiting value that $\frac{R(\boldsymbol{\tau}_{\text{homo}})}{R(\boldsymbol{\tau}^{\star})}$ converges as $\rho$ or $K$ grows large. It is not hard to see that the minimum value of $C^{\star}(\lambda)$ is $0.75$, which is achieved when $\lambda = 1$. Therefore, the maximum optimality loss arising from the use of homogenous threshold feedback policies for a two-user single beam system is $25\%$.

Up to now, we have only focused on the single beam case to explain why homogenous threshold feedback policies may sometimes become suboptimal to use. Based on the arguments above, we provide further insights as to why homogenous threshold feedback policies are always optimal to use when $M > 1$ for the Rayleigh fading channel model. We first note that, in contrast to the single beam case, Theorem 8 indicates a potential phase transition phenomenon in the behavior of the sum-rate in which homogenous threshold feedback policies suddenly become always optimal to use when we go from the single beam case to the multiple beams case. The main reason behind this phenomenon is the inter-beam interference when multiple beams are used to communicate with multiple MUs simultaneously. Such a multiuser operation makes the network interference limited, rather than being noise limited, when compared to the single beam case. More specifically, an increase in $\mathrm{SNR}$ implies a corresponding increase in the inter-beam interference experienced by other beams, and the system ends up operating always in the low $\mathsf{SNR}$ regime effectively when $M > 1$. Therefore, the low $\mathsf{SNR}$ Rayleigh fading behavior kicks in, and homogenous threshold feedback policies always become optimal to use. On the other hand, received signal powers improve linearly with $\mathsf{SNR}$ in the single beam case, which makes homogenous threshold feedback policies suboptimal to use in the high $\mathsf{SNR}$ regime.

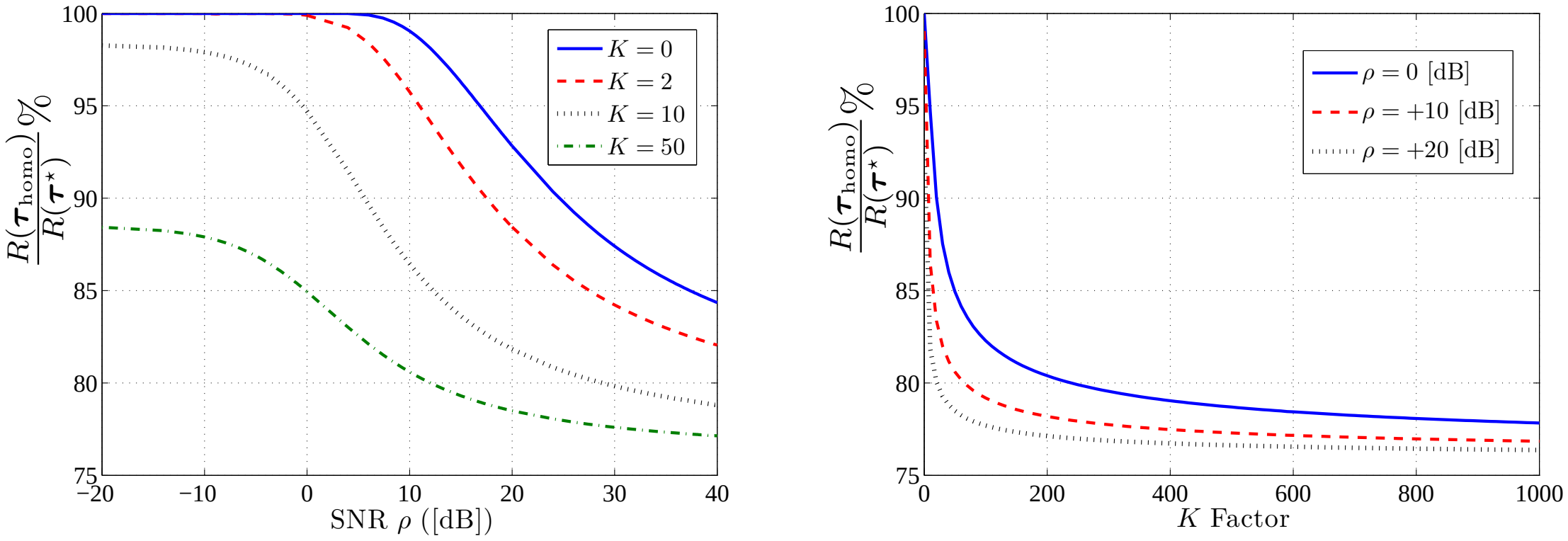

(a) As a function of $\rho$.

(b) As a function of $K$.

Fig. 11. The change of the ratio between the sum-rates achieved by homogenous and optimal threshold feedback policies as a function of $\rho$ and $K$. ($\lambda = 1$)

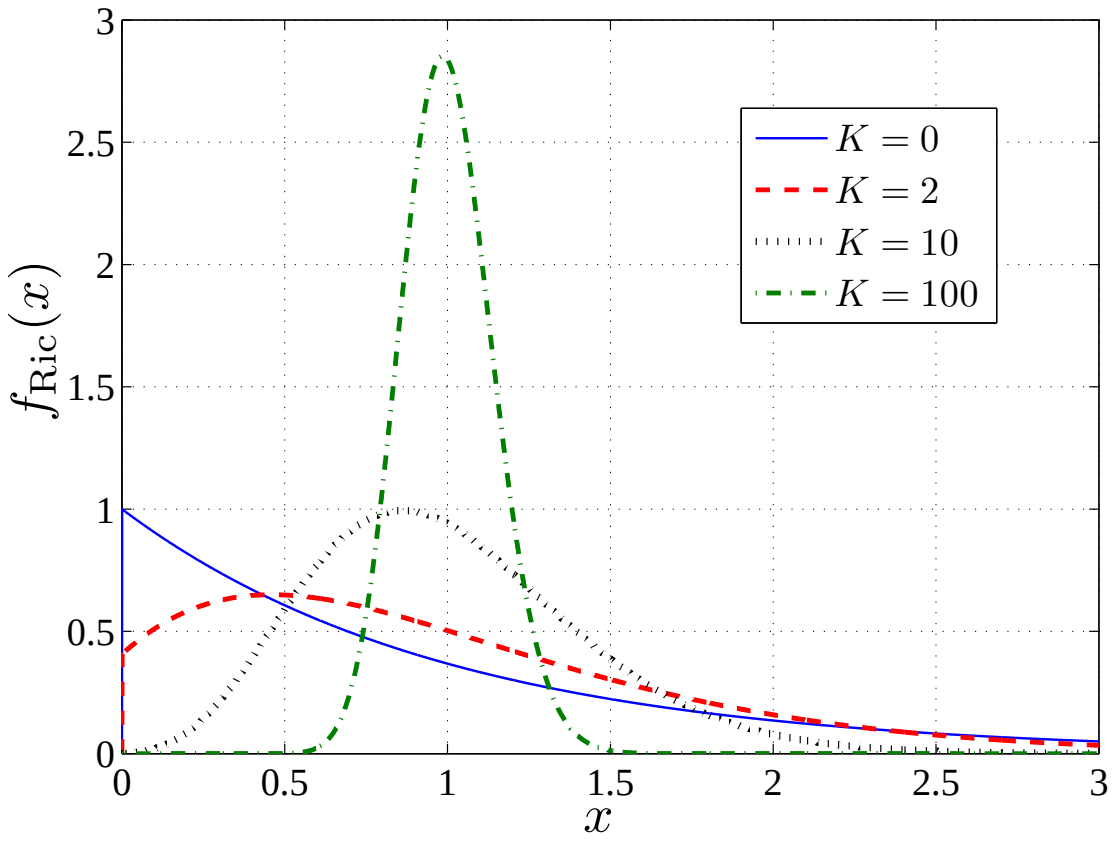

Fig. 12. Dynamic range of the SINR distribution for the Rician fading channel model for different values of $K$. ($\rho = 1$)

Although this intuition works for the Rayleigh fading channel model, it is too optimistic to ask for the optimality of homogenous threshold feedback policies for other channel models as well when $M > 1$. As our discussion above makes it apparent, the power gain due to multiuser diversity strongly depends on the parameters of the fading process determining the dynamic range of the resulting $\mathsf{SINR}$ distribution. There is no power gain to benefit from multiuser diversity by giving all MUs equal chances of channel access if the $\mathsf{SINR}$ distribution becomes increasingly more deterministic. In these instances, it is expected that a heterogenous threshold assignment will maximize the sum-rate even if the network is interference limited due to multi-beam operation. It is a potential future research interest to investigate the conditions on the parameters of the fading process to guarantee the optimality of homogenous threshold feedback policies for channel models other than the Rayleigh fading model such as Rician and Nakagami fading channels.

## VII. Conclusions

Opportunistic beamforming is an important communication strategy achieving the full CSI sum-rate capacity for vector broadcast channels to a first order by only requiring partial CSI at the BS. Nevertheless, it cannot eliminate the linear growth in the feedback load with increasing numbers of MUs in the network unless a selective feedback policy is implemented for user selection. In this paper, we have been motivated by these considerations to analyze the resulting downlink sum-rate with user selection when orthonormal beams are opportunistically allocated to MUs for the downlink communication. In particular, we have focused on the structure of optimal selective decentralized feedback policies for opportunistic beamforming under finite feedback constraints on the average number of MUs feeding back. The main findings are twofold.

We have shown that threshold feedback policies in which MUs compare their beam $\mathsf{SINR}$s with a threshold for their feedback decisions are always optimal to maximize the downlink sum-rate. This class of policies was studied in many previous works such as [13]–[18] without any formal justification for why they are the right choice for user selection. Our thresholding optimality result provides the formal justification, which holds for all fading channel models with continuous distribution functions.

Having established the optimality of threshold feedback policies, we now face an optimal threshold selection problem to maximize the sum-rate. This is a non-convex optimization problem over finite dimensional Euclidean spaces. We solve this problem by identifying an underlying Schur-concave structure in the sum-rate when it is viewed as a function of feedback probabilities. Specifically, we have obtained sufficient conditions ensuring the Schur-concavity of the sum-rate, and therefore the rate optimality of homogenous threshold feedback policies in which all MUs use the same threshold for their feedback decisions. These sufficient conditions have been provided for general fading channel models as well.

Finally, we have performed an extensive numerical and simulation study to illustrate the applications of our results to familiar fading channel models such as Rayleigh, Nakagami and Rician fading channels. With some surprise, we have shown that homogenous threshold feedback policies are not always optimal to use for general fading channels, even when all MUs experience statistically the same channel conditions. In the particular case of Rayleigh fading channels, on the other hand, homogenous threshold feedback policies have been proven to be rate-wise optimal if multiple beams are used for the downlink communication. We have also studied the optimality and sub-optimality regions for the homogenous threshold feedback policies in the Rician and Nakagami case. The detailed insights regarding when and why homogenous threshold feedback policies are rate-wise optimal or suboptimal have been provided, in conjunction with various other design and engineering perspectives.

## APPENDIX A
## LOSS EVENT AND GAIN EVENT

### A. *Proof of Lemma 1*

Set $\bar{A}_L = \left\{ \boldsymbol{\Gamma} \in \mathbb{R}_+^{M \times n} \ : \ \boldsymbol{\gamma}_1 \in \mathcal{S}_1^L \ \& \ \bar{\gamma}_1^\star < \gamma_{1,1} \right\}$. We will show $\bar{A}_L = A_L$. For all $\boldsymbol{\Gamma}$ with $\boldsymbol{\gamma}_1 \in \mathcal{S}_1^L$, MU 1 requests beam 1 under $\boldsymbol{\mathcal{F}}$, but not under $\boldsymbol{\mathcal{F}}^1$. Therefore, if $\boldsymbol{\gamma}_1 \in \mathcal{S}_1^L$ and $\bar{\gamma}_1^\star < \gamma_{1,1}$, the system using $\boldsymbol{\mathcal{F}}$ schedules MU 1 for communication along beam 1, and the system using $\boldsymbol{\mathcal{F}}^1$ schedules another MU having $\bar{\gamma}_1^\star < \gamma_{1,1}$ for communication along beam 1. This means $r^1\left(\boldsymbol{\mathcal{F}}^1, \boldsymbol{\Gamma}\right) < r^1\left(\boldsymbol{\mathcal{F}}, \boldsymbol{\Gamma}\right)$ for all $\boldsymbol{\Gamma} \in \bar{A}_L$, implying $\bar{A}_L \subseteq A_L$.

Showing $A_L \subseteq \bar{A}_L$ will complete the proof. For all $\boldsymbol{\Gamma}$ with $\bar{\gamma}_1^\star \geq \gamma_{1,1}$, both feedback policies will achieve the same throughput by scheduling the MU having $\bar{\gamma}_1^\star$. Therefore, we must have $\bar{\gamma}_1^\star < \gamma_{1,1}$ on the loss event. Now, if $\boldsymbol{\gamma}_1 \notin FB_1$, MU 1 will not feed back under $\boldsymbol{\mathcal{F}}$, which implies no potential loss on beam 1. Therefore, for all $\boldsymbol{\Gamma} \in A_L$, we must have $\boldsymbol{\gamma}_1 \in FB_1$ and $\bar{\gamma}_1^\star < \gamma_{1,1}$. If $\bar{\gamma}_1^\star < \gamma_{1,1}$ and $\boldsymbol{\gamma}_1 \in \mathcal{S}_1^R$, MU 1 requests beam 1 under both feedback policies, resulting in a neutral event. This implies that $\boldsymbol{\gamma}_1 \in \mathcal{S}_1^L$ and $\bar{\gamma}_1^\star < \gamma_{1,1}$ for all $\boldsymbol{\Gamma} \in A_L$. Therefore, we also have $A_L \subseteq \bar{A}_L$, which concludes the proof.

### B. *Proof of Lemma 2*

The proof is similar to the one given for Lemma 1. Set $\bar{A}_G = \left\{ \boldsymbol{\Gamma} \in \mathbb{R}_+^{M \times n} \ : \ \boldsymbol{\gamma}_1 \in \bar{\mathcal{S}}_1^R \ \& \ \bar{\gamma}_1^\star < \gamma_{1,1} \right\}$. We first show that $\bar{A}_G \subseteq A_G$. For all $\boldsymbol{\Gamma}$ with $\boldsymbol{\gamma}_1 \in \bar{\mathcal{S}}_1^R$ and $\bar{\gamma}_1^\star < \gamma_{1,1}$, a system using $\boldsymbol{\mathcal{F}}^1$ schedules MU 1 for communication on beam 1, but a system using $\boldsymbol{\mathcal{F}}$ schedules the MU with $\bar{\gamma}_1^\star < \gamma_{1,1}$. Therefore, $r_1\left(\boldsymbol{\mathcal{F}}^1, \boldsymbol{\Gamma}\right) > r_1\left(\boldsymbol{\mathcal{F}}, \boldsymbol{\Gamma}\right)$ if $\boldsymbol{\gamma}_1 \in \bar{\mathcal{S}}_1^R$ and $\bar{\gamma}_1^\star < \gamma_{1,1}$, implying $\bar{A}_G \subseteq A_G$.

Next, observe that the neutral event occurs for all $\mathbf{\Gamma}$ with $\bar{\gamma}_1^\star \geq \gamma_{1,1}$. Therefore, we must have $\bar{\gamma}_1^\star < \gamma_{1,1}$ on the gain event. If $\gamma_{1,1} < \tau_1$, MU 1 will not feed back under $\boldsymbol{\mathcal{F}}^1$, and therefore no rate gain is achieved by switching to $\boldsymbol{\mathcal{F}}^1$. Therefore, we must have $\gamma_{1,1} \geq \tau_1$ on the gain event. If $\boldsymbol{\gamma}_1 \in \mathcal{S}_1^R$, MU 1 still feeds back under both feedback policies, which again leads to a neutral event. Therefore, for all $\mathbf{\Gamma} \in A_G$, we must have $\boldsymbol{\gamma}_1 \in \bar{\mathcal{S}}_1^R$ and $\bar{\gamma}_1^\star < \gamma_{1,1}$, which shows that $A_G \subseteq \bar{A}_G$ and completes the proof.

## APPENDIX B
## PROOF OF LEMMA 3

$\mathcal{N}_m$ and $\mathcal{N}'_m$ are given as

$$\mathcal{N}_m = \left\{ i \in \mathcal{N} : b_i^* = m \;\&\; \gamma_{b_i^*,i} \geq \tau_i \right\}$$

and

$$\mathcal{N}'_m = \left\{ i \in \mathcal{N} : \gamma_{m,i} \geq \tau_i \right\}.$$

Thus, we have $\mathcal{N}_m \subseteq \mathcal{N}'_m$. To show the other direction, take any $i \in \mathcal{N}'_m$, and a beam index $r \neq m$. Then, $|\boldsymbol{h}_i^\top \boldsymbol{q}_m|^2 > |\boldsymbol{h}_i^\top \boldsymbol{q}_r|^2$ because $\tau_i > 1$. Therefore, the following holds.

$$\begin{aligned} \gamma_{m,i} &= \frac{|\boldsymbol{h}_i^\top \boldsymbol{q}_m|^2}{\frac{1}{\rho} + \sum_{k=1,k\neq m}^{M} |\boldsymbol{h}_i^\top \boldsymbol{q}_k|^2} \\ &> \frac{|\boldsymbol{h}_i^\top \boldsymbol{q}_r|^2}{\frac{1}{\rho} + \sum_{k=1,k\neq r}^{M} |\boldsymbol{h}_i^\top \boldsymbol{q}_k|^2} = \gamma_{r,i}. \end{aligned}$$

As a result, any MU $i \in \mathcal{N}'_m$ achieves its maximum SINR at beam $m$ if $\tau_i > 1$. This implies that $b_i^* = m$ and $i \in \mathcal{N}_m$.

## APPENDIX C
## PROOF OF LEMMA 5

Assume $\tau_2 \geq \tau_1$ (*i.e.*, $\tau_1 = \tau_{\pi(1)}$ and $\tau_2 = \tau_{\pi(2)}$) for notational simplicity. Then, for a two-user system, the rate on beam 1 as a function of the thresholds is given as

$$\begin{aligned} R^1(\tau_1, \tau_2) &= F(\tau_1) \int_{\tau_2}^{\infty} \log(1+x) dF(x) + F(\tau_2) \int_{\tau_1}^{\infty} \log(1+x) dF(x) \\ &+ \mathsf{E}\left[ \log(1 + \max\{\gamma_{1,1}, \gamma_{1,2}\}) 1_{\{\gamma_{1,1} \geq \tau_1, \gamma_{1,2} \geq \tau_2\}} \right] \\ &= F(\tau_1) \int_{\tau_2}^{\infty} \log(1+x) dF(x) + F(\tau_2) \int_{\tau_1}^{\infty} \log(1+x) dF(x) \\ &+ (1 - F(\tau_1))(1 - F(\tau_2)) \mathsf{E}\left[ \log(1 + \max\{\gamma_{1,1}, \gamma_{1,2}\}) | \gamma_{1,1} \geq \tau_1, \gamma_{1,2} \geq \tau_2 \right]. \end{aligned}$$

Let $H(x) = \mathsf{Pr}\{\max\{\gamma_{1,1}, \gamma_{1,2}\} \leq x | \gamma_{1,1} \geq \tau_1, \gamma_{1,2} \geq \tau_2\}$, *i.e.*, $H(x)$ is the CDF of $\max\{\gamma_{1,1}, \gamma_{1,2}\}$ given $\gamma_{1,1} \geq \tau_1$ and $\gamma_{1,2} \geq \tau_2$. Then,

$$H(x) = \begin{cases} \frac{F(x)-F(\tau_1)}{1-F(\tau_1)} \cdot \frac{F(x)-F(\tau_2)}{1-F(\tau_2)} & \text{if } x \geq \max\{\tau_1, \tau_2\} \\ 0 & \text{if } x < \max\{\tau_1, \tau_2\} \end{cases}. \tag{28}$$

We can write $R^1(\tau_1, \tau_2)$ as

$$\begin{aligned} R^1(\tau_1, \tau_2) &= F(\tau_1) \int_{\tau_2}^{\infty} \log(1+x) dF(x) + F(\tau_2) \int_{\tau_1}^{\infty} \log(1+x) dF(x) \\ &+ (1 - F(\tau_1))(1 - F(\tau_2)) \int_{\max\{\tau_1, \tau_2\}}^{\infty} \log(1+x) dH(x), \end{aligned} \tag{29}$$

and substituting (28) in (29) leads to

$$R^1(\tau_1, \tau_2) = \int_{\tau_2}^{\infty} \log(1+x) dF^2(x) + F(\tau_2) \int_{\tau_1}^{\tau_2} \log(1+x) dF(x), \tag{30}$$

for $\tau_2 \geq \tau_1$. For $\tau_1 \geq \tau_2$, we just switch the places of $\tau_1$ and $\tau_2$ in (30). Hence, the proof is complete.

## APPENDIX D
## RATE FOR DIFFERENT VALUES OF $\bar{\gamma}^{\star}_{\mathcal{N}'}$

### A. *Proof of Lemma 6*

Let $\bar{\xi}^{\star}_{i+1,i} = \max\left\{\bar{\gamma}_{1,\pi(i+1)}, \bar{\gamma}_{1,\pi(i)}\right\}$. From (21),

$$R^1\left(\tau_{\pi(i+1)}, \tau_{\pi(i)} | \bar{\gamma}^{\star}_{\mathcal{N}'}\right) = \log\left(1 + \bar{\gamma}^{\star}_{\mathcal{N}'}\right) \mathsf{Pr}\left\{\bar{\xi}^{\star}_{i+1,i} \leq \bar{\gamma}^{\star}_{\mathcal{N}'} | \bar{\gamma}^{\star}_{\mathcal{N}'}\right\} + \mathsf{E}\left[\log\left(1 + \bar{\xi}^{\star}_{i+1,i}\right) 1_{\left\{\bar{\xi}^{\star}_{i+1,i} > \bar{\gamma}^{\star}_{\mathcal{N}'}\right\}} | \bar{\gamma}^{\star}_{\mathcal{N}'}\right]. \tag{31}$$

Let $\mathcal{A} = \left\{\bar{\xi}^{\star}_{i+1,i} \leq \bar{\gamma}^{\star}_{\mathcal{N}'}\right\}$ and $\mathcal{B} = \left\{\xi^{\star}_{i+1,i} \leq \bar{\gamma}^{\star}_{\mathcal{N}'}\right\}$. Since $\bar{\gamma}^{\star}_{\mathcal{N}'}$ is larger than $\tau_{\pi(i+1)}$, it follows that $\mathcal{A} = \mathcal{B}$. Thus, we can write $\mathsf{Pr}\left\{\bar{\xi}^{\star}_{i+1,i} \leq \bar{\gamma}^{\star}_{\mathcal{N}'} | \bar{\gamma}^{\star}_{\mathcal{N}'}\right\} = \mathsf{Pr}\left\{\xi^{\star}_{i+1,i} \leq \bar{\gamma}^{\star}_{\mathcal{N}'} | \bar{\gamma}^{\star}_{\mathcal{N}'}\right\}$ for the first term on the righthand side of (31). For the second term, we have $\bar{\xi}^{\star}_{i+1,i} = \xi^{\star}_{i+1,i}$ since $\bar{\xi}^{\star}_{i+1,i} > \bar{\gamma}^{\star}_{\mathcal{N}'} > \tau_{\pi(i+1)}$, which concludes the proof.

### B. *Proof of Lemma 7*

When $\bar{\gamma}^{\star}_{\mathcal{N}'} < \tau_{\pi(i)}$, (31) simplifies to

$$\begin{aligned} &R^1\left(\tau_{\pi(i+1)}, \tau_{\pi(i)} | \bar{\gamma}^{\star}_{\mathcal{N}'}\right) \\ &\quad = \mathsf{E}\left[\log\left(1 + \xi^{\star}_{i+1,i}\right) 1_{\{\gamma_{1,\pi(i)} > \tau_{\pi(i)}, \gamma_{1,\pi(i+1)} > \tau_{\pi(i+1)}\}}\right] + \mathsf{E}\left[\log\left(1 + \gamma_{1,\pi(i+1)}\right) 1_{\{\gamma_{1,\pi(i+1)} > \tau_{\pi(i+1)}\}}\right] F\left(\tau_{\pi(i)}\right) \\ &\qquad + \mathsf{E}\left[\log\left(1 + \gamma_{1,\pi(i)}\right) 1_{\{\gamma_{1,\pi(i)} > \tau_{\pi(i)}\}}\right] F\left(\tau_{\pi(i+1)}\right) + \log\left(1 + \bar{\gamma}^{\star}_{\mathcal{N}'}\right) F\left(\tau_{\pi(i)}\right) F\left(\tau_{\pi(i+1)}\right). \end{aligned}$$

The first three terms on the righthand side is identical to the rate expression for the two-user system in Lemma 5. Substituting the result for the two-user case completes the proof.

### C. *Proof of Lemma 8*

For $\tau_{\pi(i)} \leq \bar{\gamma}^{\star}_{\mathcal{N}'} \leq \tau_{\pi(i+1)}$, (31) simplifies to

$$\begin{aligned} R\left(\tau_{\pi(i+1)}, \tau_{\pi(i)} | \bar{\gamma}^{\star}_{\mathcal{N}'}\right) &= \log\left(1+\bar{\gamma}^{\star}_{\mathcal{N}'}\right) F\left(\tau_{\pi(i+1)}\right) F\left(\bar{\gamma}^{\star}_{\mathcal{N}'}\right) \\ &+ \mathsf{E}\left[\log\left(1+\gamma_{1,\pi(i)}\right) 1_{\left\{\gamma_{1,\pi(i)}>\bar{\gamma}^{\star}_{\mathcal{N}'}\right\}} | \bar{\gamma}^{\star}_{\mathcal{N}'}\right] F\left(\tau_{\pi(i+1)}\right) \\ &+ \mathsf{E}\left[\log\left(1+\gamma_{1,\pi(i+1)}\right) 1_{\left\{\gamma_{1,\pi(i+1)}>\tau_{\pi(i+1)}\right\}}\right] F\left(\bar{\gamma}^{\star}_{\mathcal{N}'}\right) \\ &+ \mathsf{E}\left[\log\left(1+\xi^{\star}_{i+1,i}\right) 1_{\left\{\gamma_{1,\pi(i)}>\bar{\gamma}^{\star}_{\mathcal{N}'}, \gamma_{1,\pi(i+1)}>\tau_{\pi(i+1)}\right\}} | \bar{\gamma}^{\star}_{\mathcal{N}'}\right] \end{aligned}$$

The last three terms on the righthand side can be further simplified as in Lemma 5 for the two-user system, which completes the proof.

## APPENDIX E
## PROOF OF LEMMA 12

For $M = 1$, it is easy to get $F^{-1}(x) = -\rho \log(1-x)$. For $M > 1$, we need to find the function $F^{-1}(x)$ satisfying

$$F\left(F^{-1}(x)\right) = 1 - \frac{\exp\left(-\frac{F^{-1}(x)}{\rho}\right)}{\left(1+F^{-1}(x)\right)^{M-1}} = x.$$

The following chain of implications hold.

$$\begin{aligned} & F\left(F^{-1}(x)\right) &&= x \\ \Leftrightarrow\quad & \left(\left(1+F^{-1}(x)\right)\exp\left(\frac{1+F^{-1}(x)}{(M-1)\rho}\right)\right)^{1-M} &&= \exp\left(-\frac{1}{\rho}\right)(1-x) \\ \Leftrightarrow\quad & \frac{1+F^{-1}(x)}{(M-1)\rho}\exp\left(\frac{1+F^{-1}(x)}{(M-1)\rho}\right) &&= \frac{1}{(M-1)\rho}\left(\exp\left(-\frac{1}{\rho}\right)(1-x)\right)^{\frac{1}{1-M}} \\ \Leftrightarrow\quad & F^{-1}(x) &&= -1 + (M-1)\rho W\left(\frac{\exp\left(\frac{1}{(M-1)\rho}\right)}{(M-1)\rho}(1-x)^{\frac{1}{1-M}}\right), \end{aligned}$$

which completes the proof.

## APPENDIX F
## PROOF OF THEOREM 8

By Theorem 7, it is enough to show that $f'\left(F^{-1}(x)\right) \leq -\frac{f\left(F^{-1}(x)\right)}{\left(1+F^{-1}(x)\right)}$ for all $x \in [0,1]$. To this end, let

$$g(x) = 1 + \frac{\left(1+F^{-1}(x)\right) f'\left(F^{-1}(x)\right)}{f\left(F^{-1}(x)\right)}. \tag{32}$$

To simplify $g(x)$ further, we first put $y = F^{-1}(x)$ in (32). Then

$$g(y) = 1 + \frac{1+y}{\frac{e^{-\frac{y}{\rho}}}{(y+1)^M}\left(\frac{1}{\rho}(y+1)+M-1\right)} \times$$
$$\left(\frac{(1+y)^M e^{-\frac{y}{\rho}}\frac{1}{\rho} - (1+y)^M e^{-\frac{y}{\rho}}\left(\frac{1}{\rho}(y+1)+M-1\right)\frac{1}{\rho} - M(1+y)^{M-1}\left(\frac{1}{\rho}(y+1)+M-1\right)e^{-\frac{y}{\rho}}}{(1+y)^{2M}}\right).$$

After some further simplifications, we get

$$g(y) = 1 + \frac{\frac{1}{\rho}(y+1)}{\left(\frac{1}{\rho}(y+1)+M-1\right)} - \frac{1}{\rho}(y+1) - M.$$

Using Lemma 12, we can write $y$ as $y = -1+(M-1)\rho\bar{W}(x)$, where $\bar{W}(x) = W\left(\frac{\exp\left(\frac{1}{(M-1)\rho}\right)}{(M-1)\rho}(1-x)^{\frac{1}{1-M}}\right)$. Hence, $g(x)$ can be given as

$$\begin{aligned} g(x) &= 1 + \frac{\bar{W}(x)}{\bar{W}(x)+1} - (M-1)\bar{W}(x) - M \\ &= -\frac{(M-1)\bar{W}(x)^2 + (2M-3)\bar{W}(x) + M - 1}{\bar{W}(x)+1}, \end{aligned}$$

which is always strictly negative for $M \geq 2$. This implies $f'\left(F^{-1}(x)\right) \leq -\frac{f(F^{-1}(x))}{(1+F^{-1}(x))}$ for all $x \in [0,1]$ when $M \geq 2$, which completes the proof.